%Analytic Bethe Ansatz for Fundamental 
%Representation of Yangians
%By   Atsuo Kuniba and Junji Suzuki,
%45 pages, Plain Tex

%
%--------------------------------------------------------macros
%

\def\m@th{\mathsurround=0pt}

\def\fsquare(#1,#2){
\hbox{\vrule$\hskip-0.4pt\vcenter to #1{\normalbaselines\m@th
\hrule\vfil\hbox to #1{\hfill$\scriptstyle #2$\hfill}\vfil\hrule}$\hskip-0.4pt
\vrule}}

\def\addsquare(#1,#2){\hbox{$
	\dimen1=#1 \advance\dimen1 by -0.8pt
	\vcenter to #1{\hrule height0.4pt depth0.0pt%
	\hbox to #1{%
	\vbox to \dimen1{\vss%
	\hbox to \dimen1{\hss$\scriptstyle~#2~$\hss}%
	\vss}%
	\vrule width0.4pt}%
	\hrule height0.4pt depth0.0pt}$}}

\def\Addsquare(#1,#2){\hbox{$
	\dimen1=#1 \advance\dimen1 by -0.8pt
	\vcenter to #1{\hrule height0.4pt depth0.0pt%
	\hbox to #1{%
	\vbox to \dimen1{\vss%
	\hbox to \dimen1{\hss$~#2~$\hss}%
	\vss}%
	\vrule width0.4pt}%
	\hrule height0.4pt depth0.0pt}$}}

\def\Fsquare(#1,#2){
\hbox{\vrule$\hskip-0.4pt\vcenter to #1{\normalbaselines\m@th
\hrule\vfil\hbox to #1{\hfill$#2$\hfill}\vfil\hrule}$\hskip-0.4pt
\vrule}}

\def\naga{%
	\hbox{$\vcenter to 0.4cm{\normalbaselines\m@th
	\hrule\vfil\hbox to 1.2cm{\hfill$\cdots$\hfill}\vfil\hrule}$}}

\def\Flect(#1,#2,#3){
\hbox{\vrule$\hskip-0.4pt\vcenter to #1{\normalbaselines\m@th
\hrule\vfil\hbox to #2{\hfill$#3$\hfill}\vfil\hrule}$\hskip-0.4pt
\vrule}}

\def\PFlect(#1,#2,#3){
\hbox{$\hskip-0.4pt\vcenter to #1{\normalbaselines\m@th
\vfil\hbox to #2{\hfill$#3$\hfill}\vfil}$\hskip-0.4pt}}

\dimen1=0.5cm\advance\dimen1 by -0.8pt

\def\vnaka{\normalbaselines\m@th\baselineskip0pt\offinterlineskip%
	\vrule\vbox to 0.6cm{\vskip0.5pt\hbox to \dimen1{$\hfil\vdots\hfil$}\vfil}\vrule}

\dimen2=1.5cm\advance\dimen2 by -0.8pt

\def\vnakal{\normalbaselines\m@th\baselineskip0pt\offinterlineskip%
	\vrule\vbox to 1.2cm{\vskip7pt\hbox to \dimen2{$\hfil\vdots\hfil$}\vfil}\vrule}

\def\os{\hbox{$\fsquare(0.5cm,\hbox{0})$}\vskip-0.4pt
          \hbox{$\vnaka$}\vskip-0.4pt
	         \hbox{$\fsquare(0.5cm,\hbox{0})$}\vskip-0.4pt}

\def\bazres{\overbrace{\Fsquare(0.4cm,1)\naga\Fsquare(0.4cm,1)}^{m-j} 
\hskip-0.4pt
\overbrace{\Addsquare(0.4cm,2)\naga\Fsquare(0.4cm,2)}^{j}}
\hsize=13cm
\magnification=\magstep1
%
%--------------------------------------------------------title
%
\vskip2.5cm
\centerline{{\bf Analytic Bethe Ansatz }}\par
\centerline{{\bf for }}\par
\centerline{{\bf Fundamental Representations of
Yangians}\footnote*{Short title:  Analytic Bethe ansatz}}
\vskip1.0cm \centerline{by}
\vskip1.0cm
\centerline{Atsuo Kuniba\footnote\dag{
E-mail: atsuo@hep1.c.u-tokyo.ac.jp}
and Junji Suzuki\footnote\ddag{
E-mail: jsuzuki@tansei.cc.u-tokyo.ac.jp}} 
\vskip0.3cm
\centerline{Institute of Physics, University of Tokyo}
\centerline{Komaba 3-8-1, Meguro-ku, Tokyo 153 Japan}
\vskip7.0cm
\centerline{\bf Abstract}
\vskip0.2cm
\par
We study the analytic Bethe ansatz in
solvable vertex models associated with
the Yangian $Y(X_r)$ or its quantum affine analogue
$U_q(X^{(1)}_r)$ for $X_r = B_r, C_r$ and $D_r$.
Eigenvalue formulas are 
proposed for the transfer matrices
related to all the fundamental
representations of $Y(X_r)$.
Under the Bethe ansatz equation, we explicitly prove that
they are pole-free, a crucial 
property in the ansatz.
Conjectures are also given 
on higher representation cases by
applying the $T$-system, 
the transfer matrix functional relations proposed recently.
The eigenvalues are neatly described in terms of
Yangian analogues of the semi-standard Young tableaux.
\vfill
\eject
%
%------------------------------------------------------------------main
%
\beginsection 1. Introduction

\noindent
{\bf 1.1 General remarks.}\hskip0.3cm
Among many studies on solvable lattice models,
the Bethe ansatz is one of the most successful and 
widely applied machinery.
It was invented at very dawn of the field [1] and
is still providing rich insights.
Meanwhile, original Bethe's idea has evolved into
several versions of Bethe ans{\"a}tze called with the adjectives as
`thermodynamic' [2], `algebraic' [3], `analytic' [4,5],
`functional' [6] and so forth.
These are all powerful techniques that involve some
profound aspects.
We are yet to understand their full contents,
a challenge raised on Feynman's `last blackboard' [7].
\par
In this paper we step forward to it by developing
our recent works [8,9,10,11] further.
We shall propose eigenvalue formulas for 
several transfer matrices in the models
with the Yangian symmetry [12]
or its quantum affine analogue [13,14,15].
An interesting interplay will thereby 
be exposed between 
the representation theory of these algebras
and the analytic Bethe ansatz.
Let us explain our basic setting of the problem.
\par
\noindent
{\bf 1.2 Yang-Baxter equation and transfer matrices.}\hskip0.3cm
Consider the quantum affine algebra
$U_q(X^{(1)}_r)$ [13,14] associated with
any classical simple Lie algebra $X_r$ of rank $r$.
Throughout the paper we assume that $q$ is generic.
Let $W^{(a)}_m\, (1 \le a \le r, m \in {\bf Z}_{\ge 1})$
be the irreducible finite dimensional $U_q(X^{(1)}_r)$-module
as sketched in section 2.1.
See also [16] and [8].
For $W, W^\prime \in
\{W^{(a)}_m \mid 1 \le a \le r, m \in {\bf Z}_{\ge 1}\}$,
let $R_{W, W^\prime}(u) \in \hbox{End}(W \otimes W^\prime)$
denote the quantum $R$-matrix satisfying the Yang-Baxter equation [17]
$$
R_{W, W^\prime}(u)R_{W, W^{\prime\prime}}(u+v)
R_{W^\prime, W^{\prime\prime}}(v) = 
R_{W^\prime, W^{\prime\prime}}(v)
R_{W, W^{\prime\prime}}(u+v)
R_{W, W^\prime}(u).\eqno(1.1)
$$
Here, $u,v \in {\bf C}$ denote the spectral parameters and 
$R_{W,W^\prime}(u)$ is supposed to act as identity on
$W^{\prime\prime}$, etc.
As is well known, one has a solvable vertex model on
planar square lattice by regarding the matrix elements
of the $R$-matrix as local Boltzmann weights.
For $R_{W,W^\prime}(u)$,
the vertices take ${\rm dim }W$-states 
(resp. ${\rm dim }W^\prime$-states)
on, say, horizontal (resp. vertical) edges.
The row-to-row transfer matrix 
under the periodic boundary condition is defined by 
$$
T^{(a)}_m(u) = \hbox{Tr}_{W^{(a)}_m}\bigl(
R_{W^{(a)}_m,W^{(p)}_s}(u-w_1) \cdots 
R_{W^{(a)}_m,W^{(p)}_s}(u-w_N)\bigr)
\eqno(1.2)
$$
up to an overall scalar multiple.
Here $N$ is the system size, $w_1,\ldots,w_N$ are complex
parameters representing the inhomogeneity,
$1 \le a, p \le r$ and $m, s \in {\bf Z}_{\ge 1}$.
Following the QISM terminology [3], we say that
(1.2) is the row-to-row transfer matrix with the
{\it auxiliary space} $W^{(a)}_m$ that acts on the
{\it quantum space} $\bigl(W^{(p)}_s\bigr)^{\otimes N}$.
(More precisely, $W^{(a)}_m(u)$ and 
$\otimes_{j=1}^N W^{(p)}_s(w_j)$, 
respectively.
See section 2.1.)
Note that in (1.2) we have suppressed the quantum space 
dependence on the lhs. 
Thanks to the Yang-Baxter equation (1.1), the transfer matrices
form a commuting family
$$[T^{(a)}_m(u), T^{(a^\prime)}_{m^\prime}(u^\prime)] = 0.\eqno(1.3)$$
They can be simultaneously diagonalized 
and we shall write their eigenvalues as $\Lambda^{(a)}_m(u)$,
which is also dependent on $p$ and $s$.
Our aim is to find an explicit formula for them.
So far, the full answer is known only for
$X_r = A_r$ [18,19] and $X_r = C_2$ [10]. 
In this paper we extend the results in [20,21] for
$X_r = B_r, C_r$ and $D_r$ further 
by combining the two basic ingredients,
the analytic Bethe ansatz [5]
and the transfer matrix functional relations
($T$-system) [8,9].
Our approach renders a new insight
into the base structure of 
the module $W^{(a)}_m$ and leads to
several conjectures on $\Lambda^{(a)}_m(u)$.
Below we shall illustrate our idea 
along an exposition of the analytic Bethe ansatz (section 1.3) and 
the $T$-system (section 1.4) for the simplest example $X_r = sl(2)$.
\par\noindent
{\bf 1.3 Analytic Bethe ansatz.}\hskip0.3cm
Let us concentrate on the $X_r = sl(2)$ case
in this and the next subsections.
We write $T_m(u)$ for $T^{(1)}_m(u)$, etc. since 
the rank of $sl(2)$ is 1.
Then $W_m$ denotes the $(m+1)$-dimensional 
irreducible representation of $U_q(\hat{sl}(2))$.
For simplicity, we assume that $s=1$ in (1.2).
Then $T_1(u)$ is just the 6-vertex model transfer matrix acting on
the vectors labeled by length $N$ sequences of $+$ or $-$ states.
We take the local vertex Boltzmann weights as
$R_u(\pm,\pm,\pm,\pm) = [2+u]$,
$R_u(\pm,\mp,\pm,\mp) = [u]$ and
$R_u(\pm,\mp,\mp,\pm) = [2]$, where
the local states $+$ or $-$ are ordered anti-clockwise from
the left edge of the vertex.
The function $[u]$ is defined by
$$[u] = {q^u - q^{-u} \over q-q^{-1}}.$$
The eigenvalue $\Lambda_1(u)$
is well known and given by
$$\eqalignno{
\Lambda_1(u) &= {Q(u-1)\over Q(u+1)}\phi(u+2) +
{Q(u+3)\over Q(u+1)}\phi(u),&(1.4{\rm a})\cr
Q(u) &= \prod_{j=1}^n[u-iu_j],\quad 
\phi(u) = \prod_{j=1}^N[u-w_j].&(1.4{\rm b})\cr
}$$
Here, $0 \le n \le N/2$ is the number of the 
$-$ states in the eigenvector, which is 
preserved under the action of $T_1(u)$.
$u_j \in {\bf C}$ are
any solution of the Bethe ansatz equation (BAE)
$$ -{\phi(iu_k + 1)\over \phi(iu_k - 1)} =
{Q(iu_k + 2)\over Q(iu_k - 2)}.\eqno(1.5)
$$
On the result (1.4-5), one makes a few observations.
\par\noindent (i)
The eigenvalue has the ``dressed vacuum form (DVF)", which
means the following.
The ``vacuum vector" $+,+,\ldots,+$ is the obvious 
eigenvector with the vacuum eigenvalue
$$ 
\prod_{j=1}^NR_{u-w_j}(+,+,+,+) + 
\prod_{j=1}^NR_{u-w_j}(-,+,-,+) = \phi(u+2) + \phi(u).
\eqno(1.6)
$$
Eq.(1.4) tells that general eigenvalues have a modified form
of this with the ``dress" factors $Q/Q$ which is certainly 1
when $n = 0$.
In particular, the number of the terms in 
$\Lambda_1(u)$ is the dimension
of the auxiliary space $\hbox{dim} W_1 = 2$.
\par\noindent (ii) 
The BAE (1.5) ensures that the eigenvalues are free of poles 
for finite $u$.
The apparent pole at $u = iu_k - 1$ in (1.4a) is spurious
as the residues from the two terms cancel due to (1.5).
The eigenvalues must actually be pole-free because the 
local Boltzmann weight, hence the matrix elements 
of $T_1(u)$ are so.
\par\noindent 
(iii) Properties inherited from
the asymptotic behavior in $\vert u \vert \rightarrow \infty$ and 
the first/second inversion relations of
the $R$-matrix (vertex Boltzmann weights).
For example, one has 
$\Lambda_1(u) = (-)^N\Lambda_1(-2-u)\vert_{w_j \rightarrow -w_j,
u_j \rightarrow -u_j}$ form the last property.
See also the remark after (2.12).
\par
The analytic Bethe ansatz is 
the hypothesis that the postulates (i)-(iii) 
essentially determine a function of $u$ uniquely and that
the so obtained is the actual transfer matrix eigenvalue.
As the input data, it only uses the BAE and the $R$-matrix 
(or the vacuum eigenvalue (1.6)) which should be
normalized to be an entire function of $u$.
It was formulated in [5] by extracting the idea from 
Baxter's solution of the 8-vertex model [4].
See [10,11,20,21] for other applications.
In section 2.4, we will introduce a few more conditions
than (i)-(iii) above.
\par\noindent
{\bf 1.4 Transfer matrix functional relations.}\hskip0.3cm
The transfer matrix (1.2) obeys various 
functional relations. 
For $X_r = sl(2)$ and $s=1$ in (1.2), it is known that [18,22]
$$\eqalign{
T_m(u+1)T_m(u-1) &= T_{m+1}(u)T_{m-1}(u) + 
g_m(u)\hbox{Id},\cr
g_m(u) &= \prod_{k=0}^{m-1}
\phi(u+2k-m)\phi(u+4+2k-m),\cr}
\eqno(1.7)
$$
where $m \ge 0$ and $T_0(u) = \hbox{Id}$.
Since $T_m(u)$'s can be simultaneously diagonalized,
(1.7) may be regarded as an equation for the eigenvalues
$\Lambda_m(u)$.
By using (1.4a) and $\Lambda_0(u) = 1$ as the initial condition,
it is easy to solve the recursion (1.7) to find
$$
\Lambda_m(u) = 
\bigl(\prod_{k=1}^{m-1}\phi(u+m+1-2k)\bigr)
\sum_{j=0}^m
{Q(u-m)Q(u+m+2)\phi(u+m+1-2j)\over Q(u+m-2j)Q(u+m+2-2j)},
\eqno(1.8)
$$
in agreement with [18].
To observe a representation theoretical content,
we now set  
$$
\Fsquare(0.4cm,1) =
{Q(u-1)\over Q(u+1)}\phi(u+2),\quad
\Fsquare(0.4cm,2) =
{Q(u+3)\over Q(u+1)}\phi(u),\eqno(1.9)
$$
where we assume on the lhs that 
the spectral parameter $u$ is 
implicitly attached to the single box as well.
In this notation (1.4a) reads as
$\Lambda_1(u)
= \Fsquare(0.4cm,1) + \Fsquare(0.4cm,2)$.
Moreover, the result (1.8) for general $m$ can be expressed as follows.
$$\Lambda_m(u) = \sum_{j=0}^m \bazres . \eqno(1.10)$$
Here we interpret the tableau as the product of the $m$
functions (1.9) with the spectral parameter $u$ shifted to 
$u-m+1, u-m+3, \ldots, u+m-1$ from the left to the right.
Notice that the tableaux appearing in (1.10) are exactly
the semi-standard ones that label the weight vectors
in the $(m+1)$-dimensional irreducible representation 
$W_m$ of $U_q(\hat{sl}(2))$
(plainly, the spin $m\over 2$ representation of $sl(2)$).
In this sense 
the eigenvalues $\Lambda_m(u)$
are analogues (``Yang-Baxterizations") of the characters of
the auxiliary space $W_m$, which may be natural from (1.2).
The functional relation (1.7) for $\Lambda_m(u)$ thereby
plays the role of a character identity.
\par\noindent
{\bf 1.5 General $X_r$ case.}\hskip0.3cm
Having seen the $sl(2)$ example,
an immediate question then would be, how
the ``tableau construction" of the eigenvalues as (1.10)
can be generalized to the other algebra cases.
For $X_r = A_r$, the answer has been given in [19]
for the RSOS models [23],
which essentially includes (1.10) for $r=1$.
In this case, the $U_q(A^{(1)}_r)$-module $W^{(a)}_m$
(the auxiliary space) is a $q$-analogue of the $sl(r+1)$-module
corresponding to the $a \times m$ rectangular 
Young diagram representation.
The eigenvalue $\Lambda^{(a)}_m(u)$ is constructed as in (1.10)
from the set of the usual semi-standard tableaux labeling the 
weight vectors.
\par
An interesting feature emerges for $X_r \neq A_r$ where
$U_q(X^{(1)}_r)$-module $W^{(a)}_m$ is a $q$-analogue of 
a {\it reducible} $X_r$-module in general.
Evaluation of $\Lambda^{(a)}_m(u)$ amounts to 
finding the tableau-like objects
that label the base of such $W^{(a)}_m$.
This can actually be done by 
postulating the {\it T-system},
the transfer matrix functional relations, proposed in [8].
It is a generalization of (1.7) into arbitrary $X_r$ case
and can be solved for
$\Lambda^{(a)}_m(u)$ in terms of 
$\Lambda^{(a)}_1(u+\hbox{shift})\, (1 \le a \le r)$
(and $\Lambda^{(a)}_0(u)$ = 1).
Thus one can play the following game.
\par\noindent
{\it Step 1.} Find $\Lambda^{(1)}_1(u), \ldots, \Lambda^{(r)}_1(u)$
by the analytic Bethe ansatz.
\par\noindent
{\it Step 2.} Solve the $T$-system for $\Lambda^{(a)}_m(u)$
recursively by taking the step 1 result as the initial condition.
\par\noindent
{\it Step 3.} Find such ``tableaux" that the step 2 result is
expressed, in an analogous sense to (1.10), as
$$\Lambda^{(a)}_m(u) = \sum \hbox{tableau}(u).\eqno(1.11)$$
We shall execute the above program in a number of cases for 
$X_r = B_r, C_r$ and $D_r$.
The resulting tableau label for the base of $W^{(a)}_m$ 
exhibits an interesting contrast with
those for the crystal base [24,25] concerning 
the irreducible $X_r$-modules.
We find in several cases that the DVF (1.11), hence 
the base of $W^{(a)}_m$, 
can also be labeled by semi-standard-like tableaux
obeying remarkably simple rules.

\par\noindent
{\bf 1.6 Plan of the paper.}\hskip0.3cm
In the next section, we begin by fixing our notations 
and recall the family of the modules $W^{(a)}_m$, 
the $T$-system [8] and the BAE [26,21] for models 
with $U_q(X^{(1)}_r)$ symmetry.
The Yangian case $Y(X_r)$ corresponds to a smooth
rational limit $q \rightarrow 1$ of them.
Then we discuss the analytic Bethe ansatz
and propose a few more 
hypotheses, ``dress universality", ``top term" 
and ``coupling rule".
They supplement 
(i)-(iii) in section 1.3
and work efficiently for models with general
$U_q(X^{(1)}_r)$ symmetry.
Sections 3, 4 and 5 are devoted to the cases
$X_r = C_r, B_r$ and $D_r$, respectively.
A peculiarity for the latter two algebras
is the presence of the spin representations,
whose $U_q(X^{(1)}_r)$-analogues are certainly the members
of the family
$\{W^{(a)}_m \mid 1 \le a \le r, m \in {\bf Z}_{\ge 1} \}$.
($W^{(r)}_1$ for $B_r$ and 
$W^{(r-1)}_1, W^{(r)}_1$ for $D_r$.) 
For these algebras, we introduce two kinds of
elementary boxes corresponding to the bases of the vector and the spin
representations. 
We clarify their relation reflecting 
the fact that the former representation is contained 
in a tensor product of the latter.
These features are quite similar for $B_r$ and $D_r$ cases,
hence we shall omit many details for the latter. 
Section 6 gives the summary and discussion.
\par
\beginsection 2. $T$-system, BAE and analytic Bethe ansatz

\noindent
{\bf 2.1 Modules $W^{(a)}_m$.}\hskip0.3cm
Let us fix our notations for the data from the simple Lie algebras $X_r$.
Let $\alpha_a, \omega_a (1 \le a \le r)$ and $(\cdot\mid\cdot )$
denote the simple roots, the fundamental weights and the invariant 
bilinear form on $X_r$.
We identify the Cartan subalgebra and its dual via $(\cdot\mid\cdot)$
and normalize it as $(\alpha\mid\alpha) = 2$
for $\alpha$ = long root.
Put
$$\eqalign{
t_a & = {2 \over (\alpha_a\mid\alpha_a)} \quad 1 \le a \le r,\cr
g & = \hbox{ dual Coxeter number of } X_r.\cr}
\eqno(2.1)$$
By the definition $t_a = 1, 2$ or $3$ and 
$(\omega_a \mid \alpha_b) = \delta_{a b}/t_a$.
Enumeration of the nodes $1 \le a \le r$ 
on the Dynkin diagram is same with Table I in [8].
For $X_r = B_r (r \ge 2), C_r (r \ge 2)$ and $D_r (r \ge 4)$,
(2.1) reads explicitly as
$$\eqalign{
&t_1 = \cdots = t_{r-1} = 1, t_r = 2\, \hbox{ for } B_r,\cr
&t_1 = \cdots = t_{r-1} = 2, t_r = 1\, \hbox{ for } C_r,\cr
&\forall t_a = 1\qquad\qquad\qquad\quad\quad \,\,\hbox{ for } D_r,\cr
&g = \cases{2r-1& for $B_r$\cr
            r+1&  for $C_r$\cr
            2r-2& for $D_r$\cr}.\cr}\eqno(2.2)
$$
\par
Now we recall 
the family of modules 
$\{W^{(a)}_m \mid 1 \le a \le r, m \in {\bf Z}_{\ge 1}\}$
first introduced in [16] for the Yangian $Y(X_r)$
extending the earlier examples [26].
Precisely speaking, Yangian modules carry
a spectral parameter hence the auxiliary and the quantum
spaces in (1.2) are to be understood as
$W^{(a)}_m(u)$ and $\otimes_{j=1}^N W^{(p)}_s(w_j)$, respectively.
See [27,28] and section 3.2 in [8].
Then $W^{(a)}_m(u)$ has a characterization by the Drinfel'd
polynomials [27,28] 
$\{P_a(v) \mid 1 \le a \le r \}$ as
$$P_b(v) = \cases{
(v-u+{m-1\over t_a})(v-u+{m-3\over t_a})\cdots
(v-u-{m-1\over t_a})& for $b = a$\cr
1& otherwise \cr}.\eqno(2.3)
$$
In [28], $W^{(a)}_1(u) \, (1 \le a \le r)$ is called 
the {\it fundamental representation} of $Y(X_r)$.
Viewed as a module over $X_r \subset Y(X_r)$,
$W^{(a)}_m(u)$ is reducible in general but the 
contained irreducible components are independent of $u$.
Thus we let simply $W^{(a)}_m$ denote the $X_r$-module so obtained.
Then it is known that [16]
$$\eqalignno{
C_r;&&\cr
&W^{(a)}_m \simeq \cases{
\oplus V(k_1\omega_1 + \cdots + k_a\omega_a) \quad &$1 \le a \le r-1$\cr
V(m\omega_r) \quad &$a = r$\cr},&(2.4{\rm a})\cr
B_r \hbox{ and } D_r;&&\cr
&W^{(a)}_m \simeq \oplus 
V(k_{a_0}\omega_{a_0} + k_{a_0 + 2}\omega_{a_0 + 2} + \cdots +
k_a\omega_a) \quad 1 \le a \le r^\prime,&(2.4{\rm b})\cr
&r^\prime = \cases{r & for $B_r$\cr
                   r-2 & for $D_r$\cr},\quad
a_0 \equiv a \hbox{ mod } 2, \quad a_0 = 0 \hbox{ or } 1,
&(2.4{\rm c})\cr
&W^{(a)}_m \simeq V(m\omega_a) \quad a = r-1, r
\quad \hbox{ only for } D_r.&(2.4{\rm d})\cr}
$$
Here $\omega_0 = 0$ and 
$V(\lambda)$ denotes the irreducible $X_r$-module with 
highest weight $\lambda$.
The sum in (2.4a) is taken over non-negative integers
$k_1, \ldots, k_a$ that satisfy 
$k_1 + \cdots + k_a \le m, k_j \equiv m\delta_{j a}$
mod 2 for all $ 1 \le j \le a$.
The sum in (2.4b) extends over non-negative integers
$k_{a_0}, k_{a_0 + 2}, \ldots, k_a$ obeying the constraint
$t_a(k_{a_0} + k_{a_0+2} + \cdots + k_{a-2}) + k_a = m$.
If one depicts the highest weights in 
the sum (2.4a) and (resp. (2.4b))
by Young diagrams as usual, they correspond to those
obtained from the $a \times m$ rectangular one by successively
removing $1\times 2$ and (resp. $2 \times 1$) pieces. 
\par
As mentioned in section 3.2 of [8], we assume in this paper
that there exists
a natural $q$-analogue of these modules over the quantum affine algebra
$U_q(X^{(1)}_r)$, which will also be denoted by $W^{(a)}_m$.
When referring it 
as an $X_r$-module, it means that the corresponding 
$Y(X_r)$-module in the $q \rightarrow 1$ limit has been regarded so.
\par\noindent
{\bf 2.2 $T$-system.}\hskip0.3cm
Consider the transfer matrix (1.2)
acting on the quantum space 
$\otimes_{j=1}^N W^{(p)}_s(w_j)$.
We shall reserve the letters $p$ and $s$ for this meaning 
throughout the paper.
(See also the end of section 2.4.)
In [8], a set of functional relations, the $T$-system,
was conjectured for $U_q(X^{(1)}_r)$ symmetry models
for any $X_r$.
For $X_r = B_r, C_r$ and $D_r$ they read as follows.
$$\eqalign{
B_r:\qquad\qquad\qquad\qquad\qquad &\cr
T^{(a)}_m(u-1)T^{(a)}_m(u+1) &=
T^{(a)}_{m+1}(u)T^{(a)}_{m-1}(u) +
g^{(a)}_m(u)T^{(a-1)}_m(u)T^{(a+1)}_m(u)\cr
&\qquad\qquad\qquad\qquad\qquad 1 \le a \le r-2,\cr
T^{(r-1)}_m(u-1)T^{(r-1)}_m(u+1) &=
T^{(r-1)}_{m+1}(u)T^{(r-1)}_{m-1}(u) + 
g^{(r-1)}_m(u)T^{(r-2)}_m(u)T^{(r)}_{2m}(u),\cr
T^{(r)}_{2m}(u-{1\over 2})T^{(r)}_{2m}(u+{1\over 2}) &=
T^{(r)}_{2m+1}(u)T^{(r)}_{2m-1}(u) \cr
&+ g^{(r)}_{2m}(u)T^{(r-1)}_m(u-{1\over 2})T^{(r-1)}_m(u+{1\over 2}),\cr
T^{(r)}_{2m+1}(u-{1\over 2})T^{(r)}_{2m+1}(u+{1\over 2}) &=
T^{(r)}_{2m+2}(u)T^{(r)}_{2m}(u) +
g^{(r)}_{2m+1}(u)T^{(r-1)}_m(u)T^{(r-1)}_{m+1}(u).\cr}
\eqno(2.5{\rm a})
$$
$$\eqalign{
C_r:\qquad\qquad\qquad\qquad\qquad \quad&\cr
T^{(a)}_m(u-{1\over 2})T^{(a)}_m(u+{1\over 2}) &=
T^{(a)}_{m+1}(u)T^{(a)}_{m-1}(u) +
g^{(a)}_m(u)T^{(a-1)}_m(u)T^{(a+1)}_m(u)\cr
&\qquad\qquad\qquad\qquad\qquad 1 \le a \le r-2,\cr
T^{(r-1)}_{2m}(u-{1\over 2})T^{(r-1)}_{2m}(u+{1\over 2}) &=
T^{(r-1)}_{2m+1}(u)T^{(r-1)}_{2m-1}(u) \cr
&+ g^{(r-1)}_{2m}(u)T^{(r-2)}_{2m}(u)
T^{(r)}_m(u-{1\over 2})T^{(r)}_m(u+{1\over 2}),\cr
T^{(r-1)}_{2m+1}(u-{1\over 2})T^{(r-1)}_{2m+1}(u+{1\over 2}) &=
T^{(r-1)}_{2m+2}(u)T^{(r-1)}_{2m}(u) \cr
&+ g^{(r-1)}_{2m+1}(u)T^{(r-2)}_{2m+1}(u)
T^{(r)}_m(u)T^{(r)}_{m+1}(u),\cr
T^{(r)}_m(u-1)T^{(r)}_m(u+1) &=
T^{(r)}_{m+1}(u)T^{(r)}_{m-1}(u) +
g^{(r)}_m(u)T^{(r-1)}_{2m}(u).\cr}
\eqno(2.5{\rm b})
$$
$$\eqalign{
D_r:\qquad\qquad\qquad\qquad\qquad &\cr
T^{(a)}_m(u-1)T^{(a)}_m(u+1) &=
T^{(a)}_{m+1}(u)T^{(a)}_{m-1}(u) +
g^{(a)}_m(u)T^{(a-1)}_m(u)T^{(a+1)}_m(u)\cr
&\qquad\qquad\qquad\qquad\qquad 1 \le a \le r-3,\cr
T^{(r-2)}_m(u-1)T^{(r-2)}_m(u+1) &=
T^{(r-2)}_{m+1}(u)T^{(r-2)}_{m-1}(u) \cr
&+ g^{(r-2)}_m(u)T^{(r-3)}_m(u)T^{(r-1)}_m(u)T^{(r)}_m(u),\cr
T^{(a)}_{m}(u-1)T^{(a)}_{m}(u+1) &=
T^{(a)}_{m+1}(u)T^{(a)}_{m-1}(u)
+ g^{(a)}_{m}(u)T^{(r-2)}_m(u)\quad a = r-1, r.\cr}
\eqno(2.5{\rm c})
$$
Here the subscripts of the transfer matrices in the lhs
are taken to be positive and we assume that
$T^{(0)}_m(u) = T^{(a)}_0(u) \equiv \hbox{Id}$.
$g^{(a)}_m(u)$ is a scalar function
that depends on $W^{(p)}_s$ and satisfies
$$g^{(a)}_m(u-{1\over t_a})g^{(a)}_m(u+{1\over t_a})
= g^{(a)}_{m+1}(u)g^{(a)}_{m-1}(u).\eqno(2.6)$$
See eq.(3.18) in [8].
We have slightly changed the convention from [8]
so that $T^{(a)}_m(u+j)$ there
corresponds to $T^{(a)}_m(u+2j)$ here, etc.
A wealth of consistency for the $T$-system 
have been observed in [8,9,10,11]
for any $X_r$ and we shall assume (2.5) henceforth.
Owing to the commutativity (1.3),
one can regard (2.5) as the functional relations
on the eigenvalues $\Lambda^{(a)}_m(u)$.
($\Lambda^{(0)}_m(u) = \Lambda^{(a)}_0(u) = 1$.)
Then it can be recursively solved for
$\Lambda^{(a)}_m(u)$ in terms of 
$\Lambda^{(1)}_1(u+\hbox{shift}),\ldots,
\Lambda^{(r)}_1(u+\hbox{shift})$.
In fact, $\Lambda^{(a)}_m(u)$ will be obtainable
within a {\it polynomial}
in these functions as argued in [8].
This process corresponds to the {\it Step 2} 
mentioned in section 1.5.
\par\noindent
{\bf 2.3 Bethe ansatz equation.}\hskip0.3cm
As in (1.4), the eigenvalues $\Lambda^{(a)}_m(u)$ will be 
expressed by the solutions to the BAE [26,21]:
$$-{\phi(iu^{(a)}_k + {s\over t_p}\delta_{a p})\over
    \phi(iu^{(a)}_k - {s\over t_p}\delta_{a p})} =
\prod_{b=1}^r
{Q_b(iu^{(a)}_k + (\alpha_a\mid \alpha_b))\over
 Q_b(iu^{(a)}_k - (\alpha_a\mid \alpha_b))},\eqno(2.7)
$$
where $s$ and $p$ are the labels of the quantum space 
$\otimes_{j=1}^N W^{(p)}_s(w_j)$, 
$\phi(u)$ is given in (1.4b) and $Q_a(u)$ is defined by
$$
Q_a(u) = \prod_{j=1}^{N_a}[u - iu^{(a)}_j]\quad
1 \le a \le r.\eqno(2.8)
$$
Here $N_a$ is a non-negative integer analogous to $n$ in (1.4b).
The system size $N$ in $\phi(u)$ and $N_a$
are to be taken so that
$\omega^{(p)}_s \buildrel \rm def \over =
Ns\omega_p - \sum_{a=1}^r N_a \alpha_a \in \sum_{a=1}^r
{\bf R}_{\ge 0}\omega_a$.
The BAE (2.7) is imposed on the numbers 
$\{u^{(a)}_k \mid 1 \le a \le r, 1 \le k \le N_a \}$.
In section 5, we will consider a slightly modified version of 
(2.7) that suits the diagram automorphism symmetry in $X_r = D_r$.
\par\noindent
{\bf 2.4 Empirical rules in
Analytic Bethe ansatz.}\hskip0.3cm
As in (1.4a), the functions $Q_a(u)$ and $\phi(u)$ are the constituents 
of the dress and the vacuum parts in
the analytic Bethe ansatz, respectively.
In handling the formulas like (1.10) or (1.11),
we find it convenient to specify these parts as
$dr(\hbox{tableau})$ and $vac(\hbox{tableau})$.
For example, from the first equation in (1.9) one has
$$
\Fsquare(0.4cm,1) = dr \Fsquare(0.4cm,1)\, vac \Fsquare(0.4cm,1),\quad
dr \Fsquare(0.4cm,1) = {Q(u-1)\over Q(u+1)},\quad
vac \Fsquare(0.4cm,1) = \phi(u+2).\eqno(2.9{\rm a})
$$
In general the DVF reads
$$
\Lambda^{(a)}_m(u) = \sum
{Q_{a_1}(u+x_1) \cdots Q_{a_n}(u+x_n) \over
Q_{a_1}(u+y_1) \cdots Q_{a_n}(u+y_n)}
\phi(u+z_1) \cdots \phi(u+z_k),\eqno(2.9{\rm b})
$$
in which ratios of $Q_a$'s are the dress parts and
products of $\phi$'s are the vacuum parts.
Using these notations we now introduce 
three hypotheses, ``dress universality", ``top term" and 
``coupling rule" in the analytic Bethe ansatz.
They are the properties of mathematical interest 
rendering valuable insights into the 
auxiliary space $W^{(a)}_m$ as the 
$U_q(X^{(1)}_r)$ or the Yangian modules.
Roughly speaking, the latter two are the information on the
``highest weight vector" and the ``action" of the
Chevalley-like generators.
The hypotheses have been confirmed in several examples
and we believe they should rightly be added to 
the postulates (i)-(iii) explained in section 1.3.
\par
{\it Dress universality.}\hskip0.3cm
Let $T^{(a)}_m(u)$ and
$T^{\prime (a)}_m(u)$ be the 
transfer matrices with the same auxiliary space
$W^{(a)}_m(u)$ but acting on the 
different quantum spaces 
\par
$\otimes_{k=1}^N W^{(p)}_s(w_k)$ and 
$\otimes_{k=1}^{N^\prime} W^{(p^\prime)}_{s^\prime}(w^\prime_k)$,
respectively.
Denote by $Q_a(u)$ and $Q^\prime_a(u)$ the functions (2.8)
specified from the solutions to the BAE (2.7) for these 
quantum space choices.
Suppose one got their eigenvalues in the DVFs
$$
\Lambda^{(a)}_m(u) = \sum_{j=1}^{\hbox{dim } W^{(a)}_m}
\hbox{tab}_j,\qquad
\Lambda^{\prime (a)}_m(u) = \sum_{j=1}^{\hbox{dim } W^{(a)}_m}
\hbox{tab}^\prime_j,\eqno(2.10)
$$
where $\hbox{tab}_j$ and $\hbox{tab}^\prime_j$ denote
the terms whose vacuum parts correspond 
to the same (i.e., ``$j$-th") vector
from $W^{(a)}_m$ in the trace (1.2).
Then the dress universality is stated as
$$dr (\hbox{tab}_j) = dr (\hbox{tab}^\prime_j)
\vert_{Q^\prime_a(u) \rightarrow Q_a(u)}
\quad \hbox{ for all }\, j.
\eqno(2.11)$$
Namely, the dress part is independent of the 
quantum space choice if it is expressed in terms of $Q_a(u)$.
On the contrary, one has
$vac(\hbox{tab}_j) \neq vac(\hbox{tab}^\prime_j)
\vert_{N^\prime \rightarrow N, w^\prime_k \rightarrow w_k}$
in general if $(p^\prime, s^\prime) \neq (p,s)$.
\par
{\it Top term.}\hskip0.3cm
Among the $\hbox{dim } W^{(a)}_m$ terms in (2.10),
let $\hbox{tab}_1$ denote 
the one corresponding to the 
``highest weight vector" in $W^{(a)}_m$.
By this we mean more precisely the unique vector 
of weight $m\omega_a$ when $W^{(a)}_m$ is regarded as an
$X_r$-module in the sense of section 2.1.
Plainly, $\hbox{tab}_1$ is the analogue of the first term on the 
rhs of (1.4a).
Then the top term hypothesis reads
$$dr(\hbox{tab}_1) = {Q_a(u-{m\over t_a})\over
                      Q_a(u+{m\over t_a})}\eqno(2.12)
$$
in (2.10), which is certainly consistent to the dress universality.
It follows from (2.12) that
$$\Lambda^{(a)}_m(u)
\bigl(\Lambda^{(a)}_m(-u)\vert_{w_j \rightarrow -w_j,
u^{(b)}_k \rightarrow -u^{(b)}_k} \bigr)
= \Phi(u)\bigl(\Phi(-u)\vert_{w_j \rightarrow -w_j}\bigr) + \cdots,
$$
where $\Phi(u) = vac(\hbox{tab}_1)$ is a product of 
$\phi$'s.
This is essentially eq.(5) in [21], which is 
a consequence of the first inversion relation of the 
relevant $R$-matrix.
\par
{\it Coupling rule.}\hskip0.3cm
Regard the auxiliary space $W^{(a)}_m$ as an $X_r$-module
in the sense of section 2.1
and let $\lambda$ be a weight
without multiplicity
$$\hbox{mult}_\lambda W^{(a)}_m = 1.\eqno(2.13)
$$
Then it makes sense to denote by
$\Fsquare(0.4cm,\lambda)$ the term in (2.10)
corresponding to the $\lambda$-weight vector from $W^{(a)}_m$.
Thus 
$\Lambda^{(a)}_m(u) = \cdots + \Fsquare(0.4cm,\lambda) + \cdots$.
Now the coupling rule is stated as follows.
\par
If $\lambda$ and $\mu$ are multiplicity-free weights such that
$\lambda - \mu = \alpha_a$, then 
$$\eqalignno{
({\rm a}) &\, \Fsquare(0.4cm,\lambda) \hbox{ and } 
\Fsquare(0.4cm,\mu) \hbox{ share common poles of the form }
1/ Q_a(u+\xi)\cr
&\, \hbox{ for a certain } \xi \hbox{ depending on }
\lambda \hbox{ and } a.&(2.14{\rm a})\cr
({\rm b}) &\, \hbox{The BAE (2.7) 
guarantees } 
Res_{u=-\xi+iu^{(a)}_k}( \Fsquare(0.4cm,\lambda) + 
\Fsquare(0.4cm,\mu) ) = 0 \hbox{ in such a way that}\cr
&\, {dr \Fsquare(0.4cm,\mu)\over 
  dr \Fsquare(0.4cm,\lambda)} = 
\prod_{b=1}^r{Q_b(u+\xi+(\alpha_a\vert \alpha_b))\over
Q_b(u+\xi-(\alpha_a\vert \alpha_b))}.&(2.14{\rm b})\cr}
$$
The hypothesis tells that for $\lambda - \mu = \alpha_a$,
spurious ``poles of color $a$" in 
$\Fsquare(0.4cm, \lambda)$ and
$\Fsquare(0.4cm,\mu)$
couple into a pair yielding zero residue in total.
This is a more detailed information than just saying that
the BAE assures pole-freeness as in (ii) in section 1.3.
To determine $\xi$ is a non-trivial task in general.
From (2.14b), (2.7) and 
$\Fsquare(0.4cm,\lambda) = 
dr \Fsquare(0.4cm,\lambda) vac \Fsquare(0.4cm,\lambda)$ etc,
one deduces 
$$
{vac \Fsquare(0.4cm,\lambda) \over
 vac \Fsquare(0.4cm,\mu) }
 = {\phi(u+ \xi + {s\over t_p}\delta_{a p})\over
    \phi(u+ \xi - {s\over t_p}\delta_{a p})}\eqno(2.15)
$$
for the vacuum parts.
The last equation in (2.14b) excludes the possibility to 
exchange $\lambda$ and $\mu$ in (2.14b) and (2.15) simultaneously,
in which case 
the BAE could also have ensured the pole-freeness.
The coupling rule is certainly valid in (1.8) and (1.9) since
$\Fsquare(0.4cm,1)$ corresponds to the weight 
$\omega_1$ and $\Fsquare(0.4cm,2)$ to 
$-\omega_1 = \omega_1 - \alpha_1$ in $sl(2)$.
We will visualize (2.14) and (2.15) as
$$
\Fsquare(0.4cm,\lambda) 
{\buildrel a \over \longrightarrow} \Fsquare(0.4cm,\nu),
$$
where the number over the arrow signifies
the color of the pole shared by the two boxes.
\par
There are two more postulates 
that embody the asymptotics and
the second inversion properties mentioned in (iii) in section 1.3.
The first one is stated as
\par
{\it Character limit.}\hskip0.3cm
As said in the end of section 1.4,
the eigenvalue $\Lambda^{(a)}_m(u)$ is a Yang-Baxterization 
($u$-dependent version) of the
character of the auxiliary space $W^{(a)}_m(u)$ viewed as
an $X_r$-module in the sense of section 2.1.
Indeed, the latter can be recovered from the former as 
$$\lim_{u \rightarrow \sigma_1\infty, 
(\vert q \vert^{\sigma_2} > 1)}
q^{\tau(\sigma_1,\sigma_2)} \Lambda^{(a)}_m(u) = 
\sum_\lambda \bigl(\hbox{mult}_\lambda W^{(a)}_m\bigr)  
q^{2\sigma_1\sigma_2(\omega^{(p)}_s \vert \lambda )}
\quad \sigma_1, \sigma_2 = \pm 1,
\eqno(2.16)
$$
where the sum extends over 
all the weights in $W^{(a)}_m$,
$q^{\tau(\sigma_1,\sigma_2)}$ is some convergence factor and
$\omega^{(p)}_s$ has been specified after (2.8).
One readily sees that (2.16) is
consistent with (2.14b) and (2.15)
by computing the asymptotics of
$\Fsquare(0.4cm,\lambda)/\Fsquare(0.4cm,\mu)$.
Eq. (2.16) is also asserting that DVFs always contain
$Q_a$ via the combination 
$Q_a(u+\cdots)/Q_a(u+\cdots)$ as in (2.9b) 
and that they are homogeneous polynomials w.r.t 
$\phi(u+\cdots)$.
Thus $k$ is common in all the terms in (2.9b).
In [8,9,29], the rhs of (2.16) was denoted by
$Q^{(a)}_m(\omega^{(p)}_s)$.
It obeys the $Q$-system,
the character identity in [16], which 
was extensively
used to formulate the conjectures on 
dilogarithm identity [29,30,8,9], 
$q$-series formula for an $X^{(1)}_r$ string function [31]
and to find the $T$-system [8]. 
The limit (2.16) is essentially eq.(12) in [21].
Now we state the second postulate.
\par
{\it Crossing symmetry.}\hskip0.3cm
Most $R$-matrices enjoy the so-called 
crossing symmetry, eq.(4) in [21], from which
the second inversion relation follows.
The eigenvalue $\Lambda^{(a)}_m(u)$ inherits 
the following property form it.
$$\Lambda^{(a)}_m(u) = (-)^{kN}\Lambda^{(a)}_m(-g-u)
\vert_{w_j \rightarrow -w_j, u^{(a)}_i \rightarrow -u^{(a)}_i}.
\eqno(2.17)
$$
Here $g$ is defined in (2.1), $k$ is the order of the DVF 
w.r.t $\phi$ as in (2.9b) and $N$ is the number of lattice sites 
entering $\phi$ via (1.4b).
This is essentially eq.(6) in [21], which we call
the crossing symmetry as well.
Note that the BAE (2.7) remains unchanged under the 
simultaneous replacement
$w_j \rightarrow -w_j$ and $u^{(a)}_k \rightarrow -u^{(a)}_k$.
In particular, if $\pm \lambda$ are multiplicity-free weights
of $W^{(a)}_m$, 
the combination
$\Fsquare(0.4cm, \lambda) + \Flect(0.4cm,0.6cm,-\lambda)$
in $\Lambda^{(a)}_m(u)$ becomes same on both sides of 
(2.17) as
$$\Flect(0.4cm,0.6cm,-\lambda) = (-)^{kN}
\Fsquare(0.4cm,\lambda)
\vert_{u \rightarrow -g-u, 
w_j \rightarrow -w_j, u^{(a)}_k \rightarrow -u^{(a)}_k}.
\eqno(2.18)
$$
From the definitions of $\phi(u)$ (1.4b) and $Q_a(u)$ (2.8),
the rhs of (2.17) is then obtained from (2.9b)
by the simultaneous replacements
$$x_i \rightarrow g-x_i,\,
  y_i \rightarrow g-y_i,\,
  z_i \rightarrow g-z_i.\eqno(2.19)
$$
\par
The dress universality, top term, coupling rule,
character limit and
crossing symmetry severely limit the possible
form of the DVF in the 
analytic Bethe ansatz.
In particular if all the weights in $W^{(a)}_m$ are
multiplicity-free, (2.12), (2.14) and (2.15)
determine the DVF for $\Lambda^{(a)}_m(u)$
completely up to an overall scalar multiple.
In such cases, one even does not need the vacuum parts
a priori hence can avoid a tedious computation of the 
$R$-matrices.
The DVFs given in the subsequent sections have actually been derived
in that manner for such cases.
Except a few cases,
it is yet to be verified if 
those DVFs with $\forall Q_a(u) = 1$ yield the actual
vacuum eigenvalues obtainable from the relevant $R$-matrix 
as in (1.6).
In a sense we have partially absorbed the postulate (i) of 
section 1.3 into (2.11)-(2.15) here, which may be viewed as a 
modification of the analytic Bethe ansatz itself.
\par
Let us include a remark before closing this section.
Suppose one has found the DVF 
when the the quantum space is 
$\otimes_{j=1}^N W^{(p)}_1(w_j)$.
Then, the one for 
$\otimes_{j=1}^N W^{(p)}_s(w_j)$
can be deduced from it by the 
replacement
$$
\phi(u) \rightarrow \phi_s(u) \buildrel \rm def \over = 
\prod_{k=1}^s\phi(u+{s+1-2k\over t_p}).\eqno(2.20)
$$
To see this one just notes that 
the lhs of (2.7) is equal to 
$-{\phi_s(iu^{(a)}_k+\delta_{a p}/t_p)
\over \phi_s(iu^{(a)}_k-\delta_{a p}/t_p)}$.
See also (2.15).
In view of this we shall 
hereafter consider the $s=1$ case only with no loss of generality.
\beginsection 3. Eigenvalues for $C_r$

\noindent
{\bf 3.1 Eigenvalue $\Lambda^{(1)}_1(u)$.}\hskip0.3cm
The family of $U_q(C^{(1)}_r)$-modules 
$\{W^{(a)}_m \mid 1 \le a \le r, m \in {\bf Z}_{\ge 1} \}$
is generated by decomposing tensor products of 
$W^{(1)}_1$ as suggested in [8].
In terms of the eigenvalues, it implies that
all the $\Lambda^{(a)}_m(u)$ are contained in a suitable
product $\prod_k\Lambda_1^{(1)}(u+c_k)$.
Thus we first do the analytic Bethe ansatz for the fundamental
eigenvalue $\Lambda^{(1)}_1(u)$.
The relevant auxiliary space is $W^{(1)}_1 \simeq V(\omega_1)$ 
as an $C_r$-module from (2.4a), which is the vector 
representation.
Then all the weights are multiplicity-free and
one can apply the coupling rule (2.14).
To be concrete, we introduce the orthogonal vectors
$\epsilon_a, 1 \le a \le r$ normalized as 
$(\epsilon_a \mid \epsilon_b) = \delta_{a b}/2$
and realize the root system as follows.
$$\eqalign{
\alpha_a &= \cases{\epsilon_a - \epsilon_{a+1} &
for $1 \le a \le r-1$\cr
2\epsilon_r & for $a = r$\cr},\cr
\omega_a &= \epsilon_1 + \cdots + \epsilon_a.\cr}
\eqno(3.1)$$
Then the weights in $V(\omega_1)$ are 
$\epsilon_a$ and $-\epsilon_a (1 \le a \le r)$,
which we will abbreviate to $a$ and ${\bar a}$, respectively.
In this notation the set of weights reads
$$J = \{1, 2, \ldots, r, {\bar r},\ldots, {\bar 2}, {\bar 1}\}.
\eqno(3.2)
$$
Starting from the top term (2.12), one successively applies the 
coupling rule (2.14) to find the DVF
$$\Lambda^{(1)}_1(u) = \sum_{a \in J}\Fsquare(0.4cm,a),
\eqno(3.3)$$
with the elementary boxes defined by
$$\eqalign{
\Fsquare(0.4cm,a) &= \psi_a(u)
{Q_{a-1}(u+{a+1 \over 2})Q_a(u+{a-2 \over 2})\over
Q_{a-1}(u+{a-1 \over 2})Q_a(u+{a \over 2})}\quad 1 \le a \le r-1,\cr
\Fsquare(0.4cm,r) &= \psi_r(u)
{Q_{r-1}(u+{r+1 \over 2})Q_r(u+{r-3 \over 2})\over
Q_{r-1}(u+{r-1 \over 2})Q_r(u+{r+1 \over 2})},\cr
\Fsquare(0.4cm,{\bar r}) &= \psi_{\bar r}(u)
{Q_{r-1}(u+{r+1 \over 2})Q_r(u+{r+5 \over 2})\over
Q_{r-1}(u+{r+3 \over 2})Q_r(u+{r+1 \over 2})},\cr
\Fsquare(0.4cm,{\bar a}) &= \psi_{\bar a}(u)
{Q_{a-1}(u+{2r-a+1 \over 2})Q_a(u+{2r-a+4 \over 2})\over
Q_{a-1}(u+{2r-a+3 \over 2})Q_a(u+{2r-a+2 \over 2})}
\quad 1 \le a \le r-1,\cr}\eqno(3.4{\rm a})
$$
where we have set $Q_0(u) = 1$.
In the above, the vacuum part $\psi_a(u) = vac \Fsquare(0.4cm,a)$
is given by
$$
\psi_a(u) = \cases{
\phi(u+{p+1 \over 2})\phi(u+{2r-p+3 \over 2}) &$1 \preceq a \preceq p$\cr
\phi(u+{p-1 \over 2})\phi(u+{2r-p+3 \over 2}) 
&$p+1 \preceq a \preceq \overline{p+1}$\cr
\phi(u+{p-1 \over 2})\phi(u+{2r-p+1 \over 2}) 
&${\bar p} \preceq a \preceq {\bar 1}$\cr}
\eqno(3.4{\rm b})
$$
depending on the quantum space 
$\otimes_{j=1}^N W^{(p)}_1(w_j)$.
The symbol $\prec$ here stands for 
a total order in the set $J$ specified as
$$
1 \prec 2 \prec \cdots \prec r \prec {\bar r} \prec \cdots \prec
{\bar 2} \prec {\bar 1}.\eqno(3.5)
$$
When $p=r$, the second possibility in (3.4b) is absent.
The case $p=1$ was obtained in [21].
Note that $\Fsquare(0.4cm, 1)$ is the top term (2.12).
By the construction,
$p$ enters only the vacuum parts (3.4b) 
hence the dress universality (2.11) is valid.
The crossing symmetry (2.18) holds
between $\Fsquare(0.4cm,{\bar a})$ and $\Fsquare(0.4cm,a)$.
Under the BAE (2.7), (3.3) is pole-free because 
the coupling rule (2.14) and (2.15) as follows.
$$\eqalignno{
&Res_{u=-{b\over 2} + iu^{(b)}_k}
(\Fsquare(0.4cm,b) + \Flect(0.4cm,1.0cm,b+1)) = 0\quad 1 \le b \le r-1,
&(3.6{\rm a})\cr
&Res_{u=-{r+1\over 2} + iu^{(r)}_k}
(\Fsquare(0.4cm,r) + \Fsquare(0.4cm,{\bar r})) = 0,
&(3.6{\rm b})\cr
&Res_{u=-{2r-b+2\over 2} + iu^{(b)}_k}
(\Flect(0.4cm,1.0cm,\overline{b+1}) + \Fsquare(0.4cm,{\bar b})) = 0
\quad 1 \le b \le r-1.
&(3.6{\rm c})\cr}
$$
Following section 2.4, this can be summarized in the diagram
$$ 
\Fsquare(0.4cm,1) {\buildrel 1 \over \longrightarrow} \Fsquare(0.4cm,2)
{\buildrel 2 \over \longrightarrow} \cdots 
{\buildrel r-1 \over \longrightarrow} \Fsquare(0.4cm,r)
{\buildrel r \over \longrightarrow} \Fsquare(0.4cm,\bar{r})
{\buildrel {r-1} \over \longrightarrow}\cdots
{\buildrel 2 \over \longrightarrow}
\Fsquare(0.4cm,\bar{2}) 
{\buildrel 1 \over \longrightarrow} 
\Fsquare(0.4cm,\bar{1}) 
$$
This turns out to be identical with the crystal graph [24,25].
\par\noindent
{\bf 3.2 Eigenvalue $\Lambda^{(a)}_1(u)$.}\hskip0.3cm
Let us proceed to $\Lambda^{(a)}_1(u)$, which can be constructed 
from the elementary boxes (3.4).
For $1 \le a \le r$, let ${\cal T}^{(a)}_1$ be the set
of the tableaux of the form
$$
\hbox{
  \m@th\baselineskip0pt\offinterlineskip
   \vbox{ 
      \hbox{$\fsquare(0.5cm,\hbox{$i_1$})$}\vskip-0.4pt
      \hbox{$\vnaka$}\vskip-0.4pt
	     \hbox{$\fsquare(0.5cm,\hbox{$i_a$})$}\vskip-0.4pt
        }
      }
\eqno(3.7{\rm a})
$$
with entries $i_k \in J$ obeying the following conditions
$$\eqalignno{
&1 \preceq i_1 \prec i_2 \prec \cdots \prec i_a
\preceq {\bar 1},&(3.7{\rm b})\cr
&\hbox{If } i_k = c \hbox{ and } i_l = {\bar c}, 
\hbox{ then } r + k - l \ge c.&(3.7{\rm c})\cr}
$$
We remark that these constraints are very similar but 
different from the crystal base [24,25], where 
(3.7c) is replaced by $a+1+k-l \le c$.
We identify each element (3.7a) of ${\cal T}^{(a)}_1$ with
the product of (3.4) with the spectral parameters
$u+{a-1\over 2}, u+{a-3\over 2}, \ldots, u-{a-1\over 2}$
from the top to the bottom, namely,
$$\prod_{k=1}^a \Fsquare(0.4cm,i_k)
\vert_{u \rightarrow u + {a+1-2k\over 2}}. \eqno(3.8)
$$
Then the analytic Bethe ansatz yields the following DVF.
$$
\Lambda^{(a)}_1(u) = \sum_{T \in {\cal T}^{(a)}_1} T\quad
\qquad 1 \le a \le r,
\eqno(3.9)
$$
which reduces to (3.3) when $a=1$.
Let us observe consistency of this result before
proving that it is pole-free in section 3.3.
Firstly, the dress part of
$$
\hbox{
  \m@th\baselineskip0pt\offinterlineskip
   \vbox{ 
      \hbox{$\fsquare(0.5cm,\hbox{$1$})$}\vskip-0.4pt
      \hbox{$\vnaka$}\vskip-0.4pt
	     \hbox{$\fsquare(0.5cm,\hbox{$a$})$}\vskip-0.4pt
        }
      }
$$
is $Q_a(u-{1\over t_a})/Q_a(u+{1\over t_a})$, 
telling that the above tableau indeed gives 
the top term (2.12).
Secondly, the set ${\cal T}^{(a)}_1$ is invariant
under the interchange of the two tableaux
$$
\hbox{
  \m@th\baselineskip0pt\offinterlineskip
   \vbox{ 
      \hbox{$\fsquare(0.5cm,\hbox{$i_1$})$}\vskip-0.4pt
      \hbox{$\vnaka$}\vskip-0.4pt
	     \hbox{$\fsquare(0.5cm,\hbox{$i_a$})$}\vskip-0.4pt
        }
      }
\qquad
\hbox{
  \m@th\baselineskip0pt\offinterlineskip
   \vbox{ 
      \hbox{$\fsquare(0.5cm,\hbox{$\overline{i_a}$})$}\vskip-0.4pt
      \hbox{$\vnaka$}\vskip-0.4pt
	     \hbox{$\fsquare(0.5cm,\hbox{$\overline{i_1}$})$}\vskip-0.4pt
        }
      }
$$
and the crossing symmetry (2.18) is 
valid among them.
Thirdly, the character limit (2.16) can be proved.
This is essentially done by showing 
$$
\sharp {\cal T}^{(a)}_1 = \hbox{dim }V(\omega_a) 
= {2r\choose a} - {2r\choose a-2},\eqno(3.10)
$$
which corresponds to the $q \rightarrow 1$ limit of (2.16) 
since $W^{(a)}_1 \simeq V(\omega_a)$ as a $C_r$-module by (2.4a).
We have verified (3.10) by building injections in both directions
between the sets of depth $a$ tableaux (3.7a) breaking (3.7c)
and the depth $a-2$ ones only obeying the 
constraint as (3.7b).
Once (3.10) is established, the weight counting in (2.16) for $q \neq 1$
is shown consistent with 
eq.(2.2.2) of [25] by noting that the injections are 
weight preserving and 
$\lim_{u \rightarrow \infty, \vert q \vert > 1}
q^* \Fsquare(0.4cm,a) = q^{2(\omega^{(p)}_1\vert \epsilon_a)}$
for some $*$.
\par\noindent
{\bf 3.3 Pole-freeness of $\Lambda^{(a)}_1(u)$.}\hskip0.3cm
The quantity (3.9) passes the crucial 
condition in the analytic Bethe ansatz, namely,
\proclaim Theorem 3.3.1. 
$\Lambda^{(a)}_1(u) \, (1 \le a \le r)$ 
(3.9) is free of poles provided that
the BAE (2.7) (for $s=1$) is valid.
\par\noindent
For the proof we prepare a few Lemmas.
\proclaim Lemma 3.3.2.
For $1 \le b \le r-1$, the products
$$
\hbox{
   \normalbaselines\m@th\baselineskip0pt\offinterlineskip
   \vbox{ 
      \hbox{$\Flect(0.4cm,1.0cm,\hbox{$b$})$}\vskip-0.4pt
	     \hbox{$\Flect(0.4cm,1.0cm,\hbox{$b+1$})$}\vskip-0.4pt
        }
      }
\qquad
\hbox{
   \normalbaselines\m@th\baselineskip0pt\offinterlineskip
   \vbox{ 
      \hbox{$\Flect(0.4cm,1.0cm,\hbox{$\overline{b+1}$})$}\vskip-0.4pt
	     \hbox{$\Flect(0.4cm,1.0cm,\hbox{${\bar b}$})$}\vskip-0.4pt
        }
      }\eqno(3.11)
$$
with the spectral parameter $v$ ($v-1$) for the upper (lower)
box do not involve $Q_b$ function.
\par
The proof is straightforward by using the explicit form (3.4).
It is also elementary to check
\proclaim Lemma 3.3.3.
For $1 \le b \le r-1$, put
$$\eqalignno{
\Fsquare(0.4cm,b)_v \Flect(0.4cm,1.0cm,\overline{b+1})_{v-r+b}
&= {Q_b(v+{b\over 2}-1)\over Q_b(v+{b\over 2}+1)}X_1,
&(3.12{\rm a})\cr
\Fsquare(0.4cm,b)_v \Fsquare(0.4cm,\overline{b})_{v-r+b}
&= {Q_b(v+{b\over 2}-1)Q_b(v+{b\over 2}+2)
\over Q_b(v+{b\over 2})Q_b(v+{b\over 2}+1)}X_2,
&(3.12{\rm b})\cr
\Flect(0.4cm,1.0cm,b+1)_v \Fsquare(0.4cm,\overline{b})_{v-r+b}
&= {Q_b(v+{b\over 2}+2)\over Q_b(v+{b\over 2})}X_3,
&(3.12{\rm c})\cr}
$$
where the indices specify the spectral parameters attached
to the boxes (3.4).
Then $X_i$'s do not involve $Q_b$ function.
\par
The point is that (3.12a) and (3.12c) 
have only one $Q_b$ function in their denominators after some
cancellations owing to the spectral parameter choice
$v, v-r+b$.
\proclaim Lemma 3.3.4.
For $1 \le b \le r-1$, let the tableaux
$$
\hbox{
   \normalbaselines\m@th\baselineskip0pt\offinterlineskip
   \vbox{ 
      \hbox{$\Flect(0.6cm,1.0cm,\hbox{$\xi$})$}\vskip-0.4pt
      \hbox{$\Flect(0.4cm,1.0cm,\hbox{$b$})$}\vskip-0.4pt
      \hbox{$\Flect(0.6cm,1.0cm,\hbox{$\eta$})$}\vskip-0.4pt
      \hbox{$\Flect(0.4cm,1.0cm,\hbox{$\overline{b+1}$})$}\vskip-0.4pt
      \hbox{$\Flect(0.6cm,1.0cm,\hbox{$\zeta$})$}\vskip-0.4pt
        }
      }
\quad\hbox{ or }\quad
\hbox{
   \normalbaselines\m@th\baselineskip0pt\offinterlineskip
   \vbox{ 
      \hbox{$\Flect(0.6cm,1.0cm,\hbox{$\xi$})$}\vskip-0.4pt
      \hbox{$\Flect(0.4cm,1.0cm,\hbox{$b+1$})$}\vskip-0.4pt
      \hbox{$\Flect(0.6cm,1.0cm,\hbox{$\eta$})$}\vskip-0.4pt
      \hbox{$\Flect(0.4cm,1.0cm,\hbox{$\overline{b}$})$}\vskip-0.4pt
      \hbox{$\Flect(0.6cm,1.0cm,\hbox{$\zeta$})$}\vskip-0.4pt
        }
      }
\eqno(3.13)
$$
be the elements in ${\cal T}^{(a)}_1$ such that the columns
$\Fsquare(0.4cm,\xi), \Fsquare(0.4cm,\eta)$ and
$\Fsquare(0.4cm,\zeta)$ do not contain the boxes with entries
$b,b+1,\overline{b+1}$ and $\overline{b}$.
Then the length of $\Fsquare(0.4cm,\eta)$ is less than $r-b$.
\par
\noindent
{\it Proof.}\hskip0.3cm
We shall show this with respect to the first tableau in (3.13).
The proof for the second one is similar.
Suppose on the contrary that 
the length $L$ of $\Fsquare(0.4cm,\eta)$ satisfies
$$L \ge r-b. \eqno(3.14)$$
Due to (3.7b) and (3.5), $\Fsquare(0.4cm,\eta)$
then consists of the elementary boxes with entries from
$\{b+2,b+3,\ldots,r,\overline{r},\overline{r-1},\ldots,\overline{b+2}\}$.
By (3.14), there must be at least one letter 
$c \in \{b+2,b+3,\ldots,r\}$ such that
both $\Fsquare(0.4cm,c)$ and $\Fsquare(0.4cm,\overline{c})$
occur in $\Fsquare(0.4cm,\eta)$.
Let $c_0$ be the smallest among such $c$'s.
Then $\Fsquare(0.4cm,\eta)$ has the structure
$$
\hbox{
   \normalbaselines\m@th\baselineskip0pt\offinterlineskip
   \vbox{ 
      \hbox{$\Flect(0.6cm,0.6cm,\hbox{$\eta_1$})$}\vskip-0.4pt
      \hbox{$\Flect(0.4cm,0.6cm,\hbox{$c_0$})$}\vskip-0.4pt
      \hbox{$\Flect(0.6cm,0.6cm,\hbox{$\eta_2$})$}\vskip-0.4pt
      \hbox{$\Flect(0.4cm,0.6cm,\hbox{$\overline{c_0}$})$}\vskip-0.4pt
      \hbox{$\Flect(0.6cm,0.6cm,\hbox{$\eta_3$})$}\vskip-0.4pt
        }
      }\eqno(3.15)
$$
By the definition, the columns 
$\Fsquare(0.4cm,\eta_1)$ and $\Fsquare(0.4cm,\eta_3)$
contain only boxes
$\Fsquare(0.4cm,q)$ and $\Fsquare(0.4cm,\overline{q})$
for $b+2 \le q \le c_0-1$, respectively.
Moreover,
$\Fsquare(0.4cm,q)$ and $\Fsquare(0.4cm,\overline{q})$
must not be present simultaneously.
Thus the total length $L_{13}$ of 
$\Fsquare(0.4cm,\eta_1)$ and $\Fsquare(0.4cm,\eta_3)$
should satisfy 
$$L_{13}\le (c_0-1)-(b+2)+1 = c_0-b-2.\eqno(3.16)$$
Since (3.15) is a part of the first tableau
in (3.13) belonging to ${\cal T}^{(a)}_1$,
the condition (3.7c) must be valid for the 
$c_0, \overline{c_0}$ pair.
In terms of the length $L_2$ of the column 
$\Fsquare(0.4cm,\eta_2)$, (3.7c) reads
$$r - L_2 - 1 \ge c_0.\eqno(3.17)$$
Combining (3.14), (3.16) and (3.17), we have the contradiction
$$r - b \le L = L_{13} + L_2 + 2 \le (c_0 - b - 2) +
(r - c_0 - 1) + 2 = r - b - 1,
$$
owing to the onset assumption
(3.14), and thus finish the proof.
\par
With these Lemmas, we proceed to\par\noindent
{\it Proof of Theorem 3.3.1.}\hskip0.3cm
We shall show that color $b$ singularity 
is spurious, i.e.,
\par\noindent 
$Res_{u = iu^{(b)}_k + \cdots}\Lambda^{(a)}_1(u) = 0$
for each $2 \le b \le r-1$.
The remaining cases $b=1$ and $r$ can be verified similarly and
more easily.
Among the elementary boxes (3.4a), 
the factor $1/Q_b(u+\cdots)$ enters only 
$\Fsquare(0.4cm,b), \Flect(0.4cm,1.0cm,b+1),
\Flect(0.4cm,1.0cm,\overline{b+1})$ and
$\Fsquare(0.4cm,{\bar b})$.
Thus one has to keep track of only these four 
boxes appearing in (3.7a).
Accordingly, let us write (3.9) as
$\Lambda^{(a)}_1(u) = S_0 + S_1 + \cdots + S_4$, where
$S_k$ denotes the partial sum over the tableaux (3.7a) 
containing precisely $k$ boxes among the above four.
Obviously $S_0$ is free of $1/Q_b(u+\cdots)$.
So is $S_4$ because the relevant tableaux
involve 
both of the $2\times 1$ patterns in (3.11) and therefore
do not contain $Q_b$ by Lemma 3.3.2.
Next consider $S_1$ which is the sum over the tableaux of the form
$$
\hbox{
   \normalbaselines\m@th\baselineskip0pt\offinterlineskip
   \vbox{ 
      \hbox{$\Flect(0.6cm,0.4cm,\hbox{$\xi$})$}\vskip-0.4pt
      \hbox{$\Flect(0.4cm,0.4cm,\hbox{$b$})$}\vskip-0.4pt
      \hbox{$\Flect(0.6cm,0.4cm,\hbox{$\eta$})$}\vskip-0.4pt
        }
      }
\quad
\hbox{
   \normalbaselines\m@th\baselineskip0pt\offinterlineskip
   \vbox{ 
      \hbox{$\Flect(0.6cm,1.0cm,\hbox{$\xi$})$}\vskip-0.4pt
      \hbox{$\Flect(0.4cm,1.0cm,\hbox{$b+1$})$}\vskip-0.4pt
      \hbox{$\Flect(0.6cm,1.0cm,\hbox{$\eta$})$}\vskip-0.4pt
        }
      }
\quad
\hbox{
   \normalbaselines\m@th\baselineskip0pt\offinterlineskip
   \vbox{ 
      \hbox{$\Flect(0.6cm,1.0cm,\hbox{$\xi$})$}\vskip-0.4pt
      \hbox{$\Flect(0.4cm,1.0cm,\hbox{$\overline{b+1}$})$}\vskip-0.4pt
      \hbox{$\Flect(0.6cm,1.0cm,\hbox{$\eta$})$}\vskip-0.4pt
        }
      }
\quad
\hbox{
   \normalbaselines\m@th\baselineskip0pt\offinterlineskip
   \vbox{ 
      \hbox{$\Flect(0.6cm,0.4cm,\hbox{$\xi$})$}\vskip-0.4pt
      \hbox{$\Flect(0.4cm,0.4cm,\hbox{${\bar b}$})$}\vskip-0.4pt
      \hbox{$\Flect(0.6cm,0.4cm,\hbox{$\eta$})$}\vskip-0.4pt
        }
      }
$$
Here $\Fsquare(0.4cm,\xi)$ and $\Fsquare(0.4cm,\eta)$
stand for columns with total length $a-1$ and they do not contain 
$\Fsquare(0.4cm,b),
\Flect(0.4cm,1.0cm,b+1),
\Flect(0.4cm,1.0cm,\overline{b+1})$ and 
$\Fsquare(0.4cm,\overline{b})$.
From (3.6),
color $b$ residues in the first and second (third and fourth) tableaux
sum up to zero.
By the same reason $S_3$ is free of color $b$ singularities since
the relevant tableaux must contain one of (3.11).
Thus we are left with $S_2$, whose summands are classified into 
the following four types:
$$
\hbox{
   \normalbaselines\m@th\baselineskip0pt\offinterlineskip
   \vbox{ 
      \hbox{$\Flect(0.6cm,1.0cm,\hbox{$\xi$})$}\vskip-0.4pt
      \hbox{$\Flect(0.4cm,1.0cm,\hbox{$b$})$}\vskip-0.4pt
      \hbox{$\Flect(0.6cm,1.0cm,\hbox{$\eta$})$}\vskip-0.4pt
      \hbox{$\Flect(0.4cm,1.0cm,\hbox{$\overline{b+1}$})$}\vskip-0.4pt
      \hbox{$\Flect(0.6cm,1.0cm,\hbox{$\zeta$})$}\vskip-0.4pt
        }
      }
\quad
\hbox{
   \normalbaselines\m@th\baselineskip0pt\offinterlineskip
   \vbox{ 
      \hbox{$\Flect(0.6cm,1.0cm,\hbox{$\xi$})$}\vskip-0.4pt
      \hbox{$\Flect(0.4cm,1.0cm,\hbox{$b$})$}\vskip-0.4pt
      \hbox{$\Flect(0.6cm,1.0cm,\hbox{$\eta$})$}\vskip-0.4pt
      \hbox{$\Flect(0.4cm,1.0cm,\hbox{$\overline{b}$})$}\vskip-0.4pt
      \hbox{$\Flect(0.6cm,1.0cm,\hbox{$\zeta$})$}\vskip-0.4pt
        }
      }
\quad
\hbox{
   \normalbaselines\m@th\baselineskip0pt\offinterlineskip
   \vbox{ 
      \hbox{$\Flect(0.6cm,1.0cm,\hbox{$\xi$})$}\vskip-0.4pt
      \hbox{$\Flect(0.4cm,1.0cm,\hbox{$b+1$})$}\vskip-0.4pt
      \hbox{$\Flect(0.6cm,1.0cm,\hbox{$\eta$})$}\vskip-0.4pt
      \hbox{$\Flect(0.4cm,1.0cm,\hbox{$\overline{b+1}$})$}\vskip-0.4pt
      \hbox{$\Flect(0.6cm,1.0cm,\hbox{$\zeta$})$}\vskip-0.4pt
        }
      }
\quad
\hbox{
   \normalbaselines\m@th\baselineskip0pt\offinterlineskip
   \vbox{ 
      \hbox{$\Flect(0.6cm,1.0cm,\hbox{$\xi$})$}\vskip-0.4pt
      \hbox{$\Flect(0.4cm,1.0cm,\hbox{$b+1$})$}\vskip-0.4pt
      \hbox{$\Flect(0.6cm,1.0cm,\hbox{$\eta$})$}\vskip-0.4pt
      \hbox{$\Flect(0.4cm,1.0cm,\hbox{$\overline{b}$})$}\vskip-0.4pt
      \hbox{$\Flect(0.6cm,1.0cm,\hbox{$\zeta$})$}\vskip-0.4pt
        }
      }
\eqno(3.18)
$$
Here, $\Fsquare(0.4cm,\xi), \Fsquare(0.4cm,\eta)$ and
$\Fsquare(0.4cm,\zeta)$ are columns without 
$\Fsquare(0.4cm,b),
\Flect(0.4cm,1.0cm,b+1),
\Flect(0.4cm,1.0cm,\overline{b+1})$ and 
$\Fsquare(0.4cm,\overline{b})$.
We are going to show that the sum of the four tableaux (3.18)
is free of color $b$ singularity for any fixed 
$\Fsquare(0.4cm,\xi), \Fsquare(0.4cm,\eta)$ and
$\Fsquare(0.4cm,\zeta)$.
Denoting their lengths by $k-1, l-k-1$ and $a-l$, respectively,
we consider the cases 
$r+k-l\ge b+1, r+k-l=b$ and $r+k-l\le b-1$ separately.
If $r+k-l \ge b+1$, all the four tableaux (3.18) actually belong
to ${\cal T}^{(a)}_1$ and 
the pole-freeness of their sum follows straightforwardly
from (3.6).
If $r+k-l = b$, the third tableau in (3.18) is absent 
since it breaks (3.7c).
Up to an overall factor not containing $Q_b$, 
the remaining three terms
are proportional to those in (3.12) for some $v$.
From Lemma 3.3.3,
their sum
has zero residue both at
$v=-{b\over 2}+iu^{(b)}_k$ by (3.6a) and at 
$v=-{b\over 2}-1+iu^{(b)}_k$ by (3.6c).
Finally, we consider the case $r+k-l\le b-1$, when
the second and third tableaux in (3.18) do not exist
because they both break (3.7c).
In fact, the first and the fourth ones are also absent.
This is because $r+k-l \le b-1$ is equivalent to
saying that the length of 
$\Fsquare(0.4cm,\eta)$ is not less than $r-b$
against Lemma 3.3.4.
Thus $S_2$ is free of color $b$ poles, which completes the
proof of the Theorem.
\par
\noindent
{\bf 3.4 Eigenvalue $\Lambda^{(1)}_m(u)$.}\hskip0.3cm
The result (3.9) accomplishes {\it Step} 1 in section 1.5 
already with the tableau language sought in {\it Step} 3.
The remaining task is {\it Step} 2, i.e., to find the 
eigenvalues $\Lambda^{(a)}_m(u)$ for higher $m$
by solving the $T$-system (2.5b) with
$$\eqalign{
g^{(a)}_m(u) &= 1 \qquad\hbox{ for } 1 \le a \le r-1,\cr
g^{(r)}_m(u) &= \prod_{k=1}^m g^{(r)}_1(u+m+1-2k),\cr
g^{(r)}_1(u) &= \psi_1(u+{r+1\over 2})
\psi_{\overline{1}}(u-{r+1\over 2}),\cr}
$$
under the
initial conditions $\Lambda^{(a)}_0(u) = 1$ and (3.9).
So far we have done this only partially to get a
conjecture on $\Lambda^{(1)}_m(u)$.
To present it we introduce
a set ${\cal T}^{(1)}_m$ ($m \in {\bf Z}_{\ge 1}$)
of the tableaux having the form
$$
\fsquare(0.4cm,i_1)\naga\fsquare(0.4cm,i_k)\hskip-0.4pt
\overbrace{
\addsquare(0.4cm,\overline{r})\addsquare(0.4cm,r)\naga
\fsquare(0.4cm,\overline{r})\addsquare(0.4cm,r)
}^{2n} \hskip-0.4pt
\addsquare(0.4cm,\overline{j_l})\naga
\fsquare(0.4cm,\overline{j_1})
\eqno(3.19{\rm a})
$$
with the conditions
$$\eqalignno{
&k, n, l \ge 0,\qquad k + 2n + l = m,&(3.19{\rm b})\cr
&1 \le i_1 \le i_2 \le \cdots \le i_k \le r,
\quad
1 \le j_1 \le j_2 \le \cdots \le j_l \le r.&(3.19{\rm c})\cr}
$$
Writing (3.19a) simply as
$\Fsquare(0.4cm,i_1)\naga\Fsquare(0.4cm,i_m)$ with 
$i_k \in J$ (3.2),
we identify it with the product of (3.4) with
the spectral parameters
$u-{m-1\over 2}, u-{m-3\over 2}, \ldots, 
u+{m-1\over 2}$ from the left to the right, namely,
$$\prod_{k=1}^m \Fsquare(0.4cm,i_k)
\vert_{u \rightarrow u - {m+1-2k\over 2}}.\eqno(3.20)
$$
Then we conjecture that the 
$T$-system (2.5b) with the
initial condition (3.9) leads to
$$
\Lambda^{(1)}_m(u) = \sum_{T \in {\cal T}^{(1)}_m} T
\quad\qquad m \in {\bf Z}_{\ge 1}.\eqno(3.21)
$$
This is just (3.3) when $m=1$.
We remark that (3.21) consists of the correct number of terms,
$$\sharp {\cal T}^{(1)}_m = \hbox{dim } W^{(1)}_m.\eqno(3.22)
$$
To see this, note from (2.4a) that the rhs is equal to
$$\hbox{dim } V(m\omega_1) + \hbox{dim } V((m-2)\omega_1) + \cdots
+ \cases{
\hbox{dim } V(0)& $m$ even\cr
\hbox{dim } V(\omega_1) & $m$ odd\cr}.
\eqno(3.23)
$$
On the other hand, the set ${\cal T}^{(1)}_m$ is the disjoint
union of those tableaux (3.19a) with 
$n=0, 1, 2, \ldots$.
Thus it suffices to check
$$\hbox{dim } V(m\omega_1) = 
\sharp \{ (3.19{\rm a}) \in {\cal T}^{(1)}_m \mid n = 0 \}.
\eqno(3.24)
$$
Obviously the rhs is ${m+2r-1 \choose m}$, which
agrees with the lhs calculated from Weyl's dimension formula.
\par\noindent
{\bf 3.5 $C_2$ case.}\hskip0.3cm
For $C_2$ it is possible to provide the full solution 
$\Lambda^{(1)}_m(u), \Lambda^{(2)}_m(u)$ to the 
$T$-system [10].
In terms of the tableaux, 
$\Lambda^{(1)}_m(u)$ in [10] is certainly given by (3.21)
up to an inessential overall scalar reflecting a
different convention on $\Lambda^{(a)}_0(u)$.
To present the other eigenvalue $\Lambda^{(2)}_m(u)$ there,
we introduce a set ${\cal T}^{(2)}_m$ of $2\times m$ tableaux
$$
	\normalbaselines\m@th\offinterlineskip
	\vcenter{
   \hbox{\fsquare(0.4cm,i_1)\naga \hskip-0.4pt\fsquare(0.4cm,i_m)}
	      \vskip-0.4pt
   \hbox{\fsquare(0.4cm,j_1)\naga\hskip-0.4pt\fsquare(0.4cm,j_m)}
    }\eqno(3.25{\rm a})
$$
obeying the conditions
$$\eqalignno{
&\hbox{Every column belongs to } {\cal T}^{(2)}_1 
\hbox{ (3.7) for  } C_2,&(3.25{\rm b})\cr
&i_1 \preceq \cdots \preceq i_m,\quad 
\hbox{and }\quad j_1 \preceq \cdots \preceq j_m,
&(3.25{\rm c})\cr
&\hbox{The column }
	\normalbaselines\m@th\offinterlineskip
	\vcenter{
   \hbox{\Fsquare(0.5cm,1)}
	      \vskip-0.4pt
   \hbox{\Fsquare(0.5cm,\overline{1})} }
   \hbox{ is contained at most once. }&(3.25{\rm d})\cr}
$$
We identify each element (3.25a) in ${\cal T}^{(2)}_m$ with 
the product of (3.4) with the spectral parameters 
as follows.
$$
\prod_{k=1}^m\Fsquare(0.4cm,i_k) 
\vert_{u\rightarrow u-m-{1\over 2}+2k}\,
\prod_{k=1}^m\Fsquare(0.4cm,j_k)
\vert_{u\rightarrow u-m-{3\over 2}+2k}.\eqno(3.26)
$$
Namely, the shifts increase by 2 from the left to the right,
decrease by 1 from the top to the bottom and their average is 0.
Then the result in [10] reads
$$
\Lambda^{(2)}_m(u) = \sum_{T \in {\cal T}^{(2)}_m} T.
\eqno(3.27)
$$
\beginsection 4. Eigenvalues for $B_r$

As in the $C_r$ case we first introduce elementary boxes
attached to the vector representation.
Using them as the building blocks, we will construct the DVF for 
$\Lambda^{(a)}_1(u) (1 \le a \le r-1)$ and prove its pole-freeness
under the BAE.
We also conjecture an explicit form of 
$\Lambda^{(a)}_m(u)  (1 \le a \le r-1)$
in terms of tableaux made of these boxes.
\par
Compared with the $C_r$ case, a distinct feature in $B_r$ (and $D_r$)
is the existence of the spin representation.
Any finite dimensional irreducible $B_r$-module is
generated by decomposing a tensor product of the spin representation.
Thus we introduce
another kind of elementary boxes 
attached to the spin representation.
It enables a unified description of the DVFs for
all the fundamental eigenvalues 
$\Lambda^{(a)}_1(u)  (1 \le a \le r)$.
An explicit relation will be given between the two kinds
of the elementary boxes.
\par\noindent
{\bf 4.1 Eigenvalue $\Lambda^{(1)}_1(u)$.}\par\noindent
Let $\epsilon_a, 1 \le a \le r$ be the orthonormal vectors 
$(\epsilon_a \vert \epsilon_b) = \delta_{a b}$
realizing the root system as follows.
$$\eqalign{
\alpha_a &= \cases{\epsilon_a - \epsilon_{a+1} &
for $1 \le a \le r-1$\cr
\epsilon_r & for $a = r$\cr},\cr
\omega_a &= 
\cases{
\epsilon_1 + \cdots + \epsilon_a& for $1 \le a \le r-1$\cr
{1 \over 2}(\epsilon_1 + \cdots + \epsilon_r)& for $a = r$\cr}.\cr}
\eqno(4.1)$$
The auxiliary space relevant to 
$\Lambda^{(1)}_1(u)$ is 
$W^{(1)}_1 \simeq V(\omega_1)$ as an $B_r$-module.
This is the vector representation, whose weights are 
$\epsilon_a, -\epsilon_a (1 \le a \le r)$ and $0$.
By abbreviating them to $a, \overline{a}$
and $0$, the set of weights is given by
$$
J = \{1,2,\ldots,r,0,\overline{r},\ldots,\overline{1}\}.
\eqno(4.2)
$$
All the weights are multiplicity-free, therefore
one can determine the DVF from (2.12) and (2.14).
The result reads
$$\Lambda^{(1)}_1(u) = \sum_{a \in J}\Fsquare(0.4cm,a),
\eqno(4.3)$$
which is formally the same with (3.3).
The elementary boxes here are defined by
$$\eqalign{
\Fsquare(0.5cm,a)  &= \psi_{a}(u) 
      {{Q_{a-1}( u+a+1  ) 
            Q_{a}(u+a-2 )}\over
       { Q_{a-1}(u+a-1)Q_{a}(u+a)}} 
  \qquad 1\le a \le r,\cr
\Fsquare(0.5cm, 0)  &= 
  \psi_0(u) {{Q_r(u+r-2) Q_{r}(u+r+1)}\over
   { Q_{r}(u+r)Q_{r}(u+r-1)}},  \cr
\Fsquare(0.5cm,\bar{a})  &= 
\psi_{\bar{a}}(u) 
      {{Q_{a-1}( u+2r-a-2  ) 
            Q_{a}(u+2r-a+1 )}\over
       { Q_{a-1}(u+2r-a)Q_{a}(u+2r-a-1)}} 
  \qquad 1\le a \le r,\cr
}\eqno(4.4{\rm a})$$
where we have set $Q_0(u) = 1$.
The vacuum parts $\psi_{a}(u)$ depend on the quantum space
$\otimes_{j=1}^N W^{(p)}_1(w_j)$ 
and are given by 
$$\eqalign{
\psi_a(u) =\cases{ 
\phi(u+p+{1\over t_p})\phi(u+2r-p-1+{1\over t_p}) \Phi^r_p(u)
&   for $1 \preceq a \preceq p$\cr
\phi(u+p-{1\over t_p})\phi(u+2r-p-1+{1\over t_p}) \Phi^r_p(u)
&   for $p+1 \preceq a \preceq \overline{p+1}$   \cr
\phi(u+p-{1\over t_p})\phi(u+2r-p-1-{1\over t_p}) \Phi^r_p(u)
&   for $\overline{p} \preceq a \preceq \overline{1}$ \cr}
}\eqno(4.4{\rm b})$$
where
$$\eqalign{
\Phi^r_p(u) &= \prod_{j=1}^{p-1}
\phi(u+p-2j-{1\over t_p})\phi(u+2r-p+2j-1+{1\over t_p})\cr
&= \Phi^r_p(-2r+1-u)\vert_{w_k \rightarrow -w_k}.\cr}
\eqno(4.4{\rm c})
$$
The common factor $\Phi^r_p(u)$ here will play a role in section 4.7,
where the boxes here are related to those in section 4.5.
The order $\prec$ in the set $J$ (4.2) is defined by
$$
1 \prec 2 \prec \cdots \prec r \prec 0 \prec 
\overline{r} \prec \cdots \prec \overline{2} \prec
\overline{1}. \eqno(4.5)
$$
Note the top term $\Fsquare(0.4cm,1)$ (2.12), 
the dress universality (2.11) and the crossing symmetry
(2.18) for the pairs
$\Fsquare(0.4cm,a) \leftrightarrow 
\Fsquare(0.4cm,\overline{a})$ and
$\Fsquare(0.4cm,0) \leftrightarrow \Fsquare(0.4cm,0)$.
Under the BAE (2.7), (4.3) is pole-free because
the coupling rule (2.14) and (2.15) have been embodied as
$$\eqalignno{
&Res_{u=-b + iu^{(b)}_k}
(\Fsquare(0.4cm,b) + \Flect(0.4cm,1.0cm,b+1)) = 0\quad 1 \le b \le r-1,
&(4.6{\rm a})\cr
&Res_{u=-r + iu^{(r)}_k}
(\Fsquare(0.4cm,r) + \Fsquare(0.4cm,0)) = 0,
&(4.6{\rm b})\cr
&Res_{u=-r+1+iu^{(r)}_k}
(\Fsquare(0.4cm,0) + \Fsquare(0.4cm,\overline{r})) = 0,
&(4.6{\rm c})\cr
&Res_{u=-2r+b+1 + iu^{(b)}_k}
(\Flect(0.4cm,1.0cm,\overline{b+1}) + \Fsquare(0.4cm,{\bar b})) = 0
\quad 1 \le b \le r-1.
&(4.6{\rm d})\cr}
$$
Thus we have a diagram
$$
\Fsquare(0.4cm,1) {\buildrel 1 \over \longrightarrow} \Fsquare(0.4cm,2)
{\buildrel 2 \over \longrightarrow} \cdots 
{\buildrel r-1 \over \longrightarrow} \Fsquare(0.4cm,r)
{\buildrel r \over \longrightarrow} \Fsquare(0.4cm,0) 
{\buildrel r \over \longrightarrow} \Fsquare(0.4cm,\bar{r})
{\buildrel {r-1} \over \longrightarrow} \cdots
{\buildrel 2 \over \longrightarrow}
\Fsquare(0.4cm,\bar{2}) 
{\buildrel 1 \over \longrightarrow} 
\Fsquare(0.4cm,\bar{1}) 
$$
This is again identical with the crystal graph [24,25].
For $p=1$, (4.3) has been known earlier in [21].
\par\noindent
{\bf 4.2 Eigenvalue $\Lambda^{(a)}_1(u)$ for $1 \le a \le r-1$.}
\hskip0.3cm
For $1 \le a \le r-1$, let ${\cal T}^{(a)}_1$ be the set of
the tableaux of the form (3.7a) with $i_k \in J$ (4.2) 
obeying the condition
$$
i_k \prec i_{k+1} \hbox{ or } i_k = i_{k+1} = 0
\,\,\hbox{ for any } 1 \le k \le a-1.\eqno(4.7)
$$
Namely, the entries must increase strictly from the top to the bottom
in the sense of (4.5) except a possible segment of consecutive $0$'s.
We identify each element (3.7a) of ${\cal T}^{(a)}_1$ with
the product of (4.4a) with the spectral parameters
$u+a-1, u+a-3,\ldots,u-a+1$ from the top to the bottom
$$
\prod_{k=1}^a \Fsquare(0.4cm,i_k)
\vert_{u \rightarrow u+a+1-2k}.\eqno(4.8)
$$
Then the analytic Bethe ansatz yields the following DVF.
$$
\Lambda^{(a)}_1(u) = 
{1\over F^{(p,r)}_a(u)} 
\sum_{T \in {\cal T}^{(a)}_1} T\quad
\qquad 1 \le a \le r-1,\eqno(4.9{\rm a})
$$
where the scalar $F^{(p,r)}_a(u)$ is defined by
$$\eqalign{
&F^{(p,r)}_a(u) \cr
&= \prod_{j=1}^{a-1}\prod_{k=0}^{p-1}
\phi(u+p+a-1-{1\over t_p}-2j-2k)
\phi(u+2r-p-a+{1\over t_p}+2j+2k)\cr
&= F^{(p,r)}_a(-2r+1-u)\vert_{w_k \rightarrow -w_k}.\cr}
\eqno(4.9{\rm b})
$$
Notice that $F^{(p,r)}_1(u) = 1$ hence (4.9a)
reduces to (4.3) when $a=1$.
From (4.4c) and (4.27c) in section 4.5, (4.9b) can also be written as
$$\eqalignno{
F^{(p,r)}_a(u) &=
\prod_{j=1}^{a-1}\phi(u+p+a-1-{1\over t_p}-2j)
\phi(u+2r+p+a-2+{1\over t_p}-2j)\cr
&\times\prod_{j=1}^{a-1}\Phi^r_p(u+a-1-2j)
&(4.10{\rm a})\cr
&= \prod_{j=1}^{a-1}
\psi^{(p,r)}_0(u+r-a-{1\over 2}+2j)
\psi^{(p,r)}_p(u-r+a+{1\over 2}-2j).
&(4.10{\rm b})\cr}
$$
By using (4.10a), it can be checked that
each summand $T$ in (4.9a) contains the factor 
$F^{(p,r)}_a(u)$ and $\Lambda^{(a)}_1(u)$ is homogeneous
of order $2p$ w.r.t $\phi(u+\cdots)$.
This will be seen more manifestly 
in Theorem 4.7.1.
One can observe the top term and the crossing symmetry
in the DVF (4.9a) as done after (3.9).
The character limit (2.16) is also valid.
To see this, we introduce a map $\chi$ from ${\cal T}^{(a)}_1$
to the Laurent polynomials 
${\bf C}[z_1,z^{-1}_1,\ldots,z_r,z^{-1}_r]$ by
$$
\raise 4ex \hbox{$\chi \Biggl($}
  \hbox{
  \m@th\baselineskip0pt\offinterlineskip
   \vbox{ 
      \hbox{$\fsquare(0.5cm,\hbox{$i_1$})$}\vskip-0.4pt
      \hbox{$\vnaka$}\vskip-0.4pt
	     \hbox{$\fsquare(0.5cm,\hbox{$i_a$})$}\vskip-0.4pt
        }
      }
\raise 4ex \hbox{$\Biggr) = y_{i_1} \cdots y_{i_a}$,}
\eqno(4.11{\rm a})
$$
where
$$
y_0 = 1,\, y_a = z_a, \, y_{\overline{a}} = z^{-1}_a, \, 1 \le a \le r.
\eqno(4.11{\rm b})
$$
In view of 
$\lim_{u \rightarrow \infty, \vert q \vert > 1}
q^* \Fsquare(0.4cm,a) = q^{2(\omega^{(p)}_1\vert \epsilon_a)}$ for some
$*$, this corresponds to taking the limit (2.16) of 
the element (3.7a).
Since $W^{(a)}_1 \simeq 
V(\omega_a) \oplus V(\omega_{a-2}) \oplus \cdots$ from (2.4b),
we are to show 
$$
\sum_{T \in {\cal T}^{(a)}_1} \chi(T) = 
ch V(\omega_a) + ch V(\omega_{a-2}) + \cdots,
\eqno(4.11{\rm c})
$$
for $1 \le a \le r-1$.
Here $ch V$ denotes the classical character of the $B_r$-module
$V$ on letters $z_1, \ldots, z_r$.
This can be easily proved from (4.7) and the known formula
$$
ch V(\omega_a) = 
\sum_{\scriptstyle i_1, \ldots, i_a \in J \atop 
       \scriptstyle i_1 \prec \cdots \prec i_a}
       y_{i_1} \cdots y_{i_a},\eqno(4.11{\rm d})
$$
for $1 \le a \le r-1$.
Eq.(4.11d) originates in 
$so(2r+1) \hookrightarrow gl(2r+1)$.
\par\noindent
{\bf 4.3 Pole-freeness of $\Lambda^{(a)}_1(u)$ for $1 \le a \le r-1$.}
\hskip0.3cm
The purpose of this section is to show
\proclaim Theorem 4.3.1. 
$\Lambda^{(a)}_1(u) (1 \le a \le r-1)$ (4.9) is free of poles
provided that the BAE (2.7) (for $s=1$) is valid.
\par
For the proof we need
\proclaim Lemma 4.3.2. For $n \in {\bf Z}_{\ge 0}$, put
$$\eqalignno{
\prod_{j=0}^n \Fsquare(0.4cm,0)_{v-2j} &=
{Q_r(v+r+1)Q_r(v+r-2n-2)\over
Q_r(v+r)Q_r(v+r-2n-1)}X_1,&(4.12{\rm a})\cr
\Fsquare(0.4cm,r)_v \prod_{j=1}^n \Fsquare(0.4cm,0)_{v-2j} &=
{Q_r(v+r-1)Q_r(v+r-2n-2)\over
Q_r(v+r)Q_r(v+r-2n-1)}X_2,&(4.12{\rm b})\cr
\Fsquare(0.4cm,\overline{r})_{v-2n}
\prod_{j=0}^{n-1} \Fsquare(0.4cm,0)_{v-2j} &=
{Q_r(v+r+1)Q_r(v+r-2n)\over
Q_r(v+r)Q_r(v+r-2n-1)}X_3,&(4.12{\rm c})\cr
\Fsquare(0.4cm,r)_v \Fsquare(0.4cm,\overline{r})_{v-2n}
\prod_{j=1}^{n-1} \Fsquare(0.4cm,0)_{v-2j} &=
{Q_r(v+r-1)Q_r(v+r-2n)\over
Q_r(v+r)Q_r(v+r-2n-1)}X_4,&(4.12{\rm d})\cr}
$$
where the indices specify the spectral parameters attached
to the boxes (4.4).
Then 
$$\eqalignno{
&X_i\hbox{'s do not involve } Q_r \hbox{ function},
&(4.13{\rm a})\cr
&{Q_r(v+r\pm 1)\over Q_r(v+r)} \hbox{ comes from the box }
\Fsquare(0.4cm,\ast)_v,
&(4.13{\rm b})\cr
&{Q_r(v+r-2n-1\pm 1)\over Q_r(v+r-2n-1)} \hbox{ comes from the box }
\Fsquare(0.4cm,\ast)_{v-2n},
&(4.13{\rm c})\cr}
$$
where $\ast = r, \overline{r}$ or $0$.
\par
This can be verified by a direct calculation.
\proclaim Lemma 4.3.3. If the BAE (2.7) ($s=1$) is valid, then
$$\eqalign{
&Res_{v=-r+iu^{(r)}_k}((4.12{\rm a}) + (4.12{\rm b})) =
 Res_{v=-r+iu^{(r)}_k}((4.12{\rm c}) + (4.12{\rm d})) = 0,\cr
&Res_{v=-r+2n+1+iu^{(r)}_k}((4.12{\rm a}) + (4.12{\rm c})) \cr
&\quad = Res_{v=-r+2n+1+iu^{(r)}_k}((4.12{\rm b}) + (4.12{\rm d})) = 0.\cr}
\eqno(4.14)
$$
\par
This follows from (4.13b,c) and (4.6b,c).
Now we proceed to
\par\noindent
{\it Proof of Theorem 4.3.1.}\hskip0.3cm
As remarked after (4.10), there is no pole originated from
the overall scalar $1/F^{(p,r)}_a(u)$ in (4.9a).
Thus one has only to show that the apparent color $b$ poles
$1/Q_b(u+\cdots)$ in 
$\sum_{T \in {\cal T}^{(a)}_1} T$ 
are spurious for all $1\le b \le r$
under the BAE.
For $1\le b \le r-1$, this can be done similarly to
the proof of Theorem 3.3.1.
In fact the present case is much easier since (4.7) is
so compared with (3.7b,c).
Henceforth we focus on the $b=r$ case which needs a separate consideration.
From (4.4a), we have only to keep track of the boxes
$\Fsquare(0.4cm,r), \Fsquare(0.4cm,0)$ and 
$\Fsquare(0.4cm,\overline{r})$ containing $Q_r$.
Let us classify the tableaux (3.7a) in ${\cal T}^{(a)}_1$ (4.7)
into the sectors labeled by
the number $n$ of $\Fsquare(0.4cm,0)$'s contained in them.
In each sector, we further divide the tableaux into four types
according to the entries $(u,d)$ in the boxes just above
and below the consecutive $\Fsquare(0.4cm,0)$'s.
$$\eqalign{
&\hbox{type } 1_n: u \neq r \hbox{ and } d \neq \overline{r},\cr
&\hbox{type } 2_n: u  =   r \hbox{ and } d \neq \overline{r},\cr
&\hbox{type } 3_n: u \neq r \hbox{ and } d  =   \overline{r},\cr
&\hbox{type } 4_n: u  =   r \hbox{ and } d  =   \overline{r}.\cr
}$$
Thus we have
$$\eqalignno{
\sum_{T \in {\cal T}^{(a)}_1} T 
&= \sum_{n=0}^a \sum_{i=1}^4 S_{n,i},&(4.15{\rm a})\cr
S_{n,i} &= \sum_{T \in \hbox{type } i_n} T,&(4.15{\rm b})\cr
S_{a,2} &= S_{a,3} = S_{a,4} = S_{a-1,4} = 0.
&(4.15{\rm c})\cr}
$$
Consider the following quartet of the tableaux of
types $1_{n+1}, 2_n, 3_n$ and $4_{n-1}$, respectively.
$$
\hbox{
  \m@th\baselineskip0pt\offinterlineskip
   \vbox{ 
      \hbox{$\Flect(0.8cm,0.5cm,\hbox{$\xi$})$}\vskip-0.4pt
      \hbox{$\fsquare(0.5cm,\hbox{$0$})$}\vskip-0.4pt
      \hbox{$\fsquare(0.5cm,\hbox{$0$})$}\vskip-0.4pt
      \hbox{$\vnaka$}\vskip-0.4pt
      \hbox{$\fsquare(0.5cm,\hbox{$0$})$}\vskip-0.4pt
      \hbox{$\fsquare(0.5cm,\hbox{$0$})$}\vskip-0.4pt
      \hbox{$\Flect(0.8cm,0.5cm,\hbox{$\eta$})$}\vskip-0.4pt
      
        }
      }
\quad\qquad
\hbox{
  \m@th\baselineskip0pt\offinterlineskip
   \vbox{ 
      \hbox{$\Flect(0.8cm,0.5cm,\hbox{$\xi$})$}\vskip-0.4pt
      \hbox{$\fsquare(0.5cm,\hbox{$r$})$}\vskip-0.4pt
      \hbox{$\fsquare(0.5cm,\hbox{$0$})$}\vskip-0.4pt
      \hbox{$\vnaka$}\vskip-0.4pt
      \hbox{$\fsquare(0.5cm,\hbox{$0$})$}\vskip-0.4pt
      \hbox{$\fsquare(0.5cm,\hbox{$0$})$}\vskip-0.4pt
      \hbox{$\Flect(0.8cm,0.5cm,\hbox{$\eta$})$}\vskip-0.4pt
      
        }
      }
\quad\qquad
\hbox{
  \m@th\baselineskip0pt\offinterlineskip
   \vbox{ 
      \hbox{$\Flect(0.8cm,0.5cm,\hbox{$\xi$})$}\vskip-0.4pt
      \hbox{$\fsquare(0.5cm,\hbox{$0$})$}\vskip-0.4pt
      \hbox{$\fsquare(0.5cm,\hbox{$0$})$}\vskip-0.4pt
      \hbox{$\vnaka$}\vskip-0.4pt
      \hbox{$\fsquare(0.5cm,\hbox{$0$})$}\vskip-0.4pt
      \hbox{$\fsquare(0.5cm,\hbox{$\overline{r}$})$}\vskip-0.4pt
      \hbox{$\Flect(0.8cm,0.5cm,\hbox{$\eta$})$}\vskip-0.4pt
      
        }
      }
\quad\qquad
\hbox{
  \m@th\baselineskip0pt\offinterlineskip
   \vbox{ 
      \hbox{$\Flect(0.8cm,0.5cm,\hbox{$\xi$})$}\vskip-0.4pt
      \hbox{$\fsquare(0.5cm,\hbox{$r$})$}\vskip-0.4pt
      \hbox{$\fsquare(0.5cm,\hbox{$0$})$}\vskip-0.4pt
      \hbox{$\vnaka$}\vskip-0.4pt
      \hbox{$\fsquare(0.5cm,\hbox{$0$})$}\vskip-0.4pt
      \hbox{$\fsquare(0.5cm,\hbox{$\overline{r}$})$}\vskip-0.4pt
      \hbox{$\Flect(0.8cm,0.5cm,\hbox{$\eta$})$}\vskip-0.4pt
      
        }
      }
\eqno(4.16)
$$
Here, $\Fsquare(0.4cm,\xi)$ and $\Fsquare(0.4cm,\eta)$
are the columns with total length $a-n-1$ and they
do not contain $\Fsquare(0.4cm,r),
\Fsquare(0.4cm,0)$ and $\Fsquare(0.4cm,\overline{r})$.
In view of (4.8) and (4.13a), the tableaux (4.16) are proportional
to the four terms (4.12) with some $v$ up to an overall factor
not containing $Q_r$.
Thus from Lemma 4.3.3, their sum is free of
color $r$ singularity.
This is true for any fixed
$\Fsquare(0.4cm,\xi)$ and $\Fsquare(0.4cm,\eta)$ such that 
the tableaux (4.16) belong to ${\cal T}^{(a)}_1$.
Therefore $S_{n+1,1} + S_{n,2} + S_{n,3} + S_{n-1,4}$
is free of color $r$ singularity for each 
$1 \le n \le a-1$.
Due to (4.15c), the remaining terms in (4.15a) are 
$S_{1,1}, S_{0,1}, S_{0,2}$ and $S_{0,3}$.
By the definition $S_{0,1}$ is independent of $Q_r$ and
it is straightforward to check that
$S_{1,1}+S_{0,2}+S_{0,3}$ is free of color $r$ singularity
by using (4.6b,c).
This establishes the Theorem.
\par\noindent
{\bf 4.4 Eigenvalue $\Lambda^{(a)}_m(u)$ for $1\le a \le r-1$.}
\hskip0.3cm
Starting from (4.9) and $\Lambda^{(r)}_1(u)$
that will be described in section 4.5, we are to solve
the $T$-system (2.5a) with
$$\eqalign{
g^{(a)}_m(u) &= 1 \qquad \hbox{ for } 2 \le a \le r,\cr
g^{(1)}_m(u) &= \prod_{k=1}^m g^{(1)}_1(u+m+1-2k),\cr
g^{(1)}_1(u) &= F^{(p,r)}_2(u),\cr}
\eqno(4.17)
$$
where the last quantity has been given in (4.9b) and (4.10).
The solution will yield a DVF
for the general eigenvalue $\Lambda^{(a)}_m(u)$.
Here we shall present the so derived conjecture for $1 \le a \le r-1$.
\par
Let ${\cal T}^{(a)}_m (1 \le a \le r-1)$ be the set of the 
$a \times m$ rectangular tableaux containing 
$\Fsquare(0.5cm,i_{j k}), \, i_{j k} \in J$ at the 
$(j,k)$ position.
$$
	\normalbaselines\m@th\offinterlineskip
	\vcenter{
   \hbox{$\Flect(0.7cm,0.7cm,\hbox{$i_{1 1}$})\hskip-0.4pt
         \Flect(0.7cm,1.5cm,\hbox{$\cdots$}) \hskip-0.4pt
         \Flect(0.7cm,0.7cm,\hbox{$i_{1 m}$})$}
    \vskip-0.4pt
   \hbox{$\Flect(1.2cm,0.7cm,\hbox{$\vdots$})\hskip-0.4pt
         \Flect(1.2cm,1.5cm,\hbox{$\ddots$}) \hskip-0.4pt
         \Flect(1.2cm,0.7cm,\hbox{$\vdots$})$
         }
     \vskip-0.4pt
   \hbox{$\Flect(0.7cm,0.7cm,\hbox{$i_{a 1}$})\hskip-0.4pt
         \Flect(0.7cm,1.5cm,\hbox{$\cdots$}) \hskip-0.4pt
         \Flect(0.7cm,0.7cm,\hbox{$i_{a m}$})$}
         }
$$
The entries are to obey the conditions
$$\eqalignno{
&i_{j k} \prec i_{j+1 k} \hbox{ or }
i_{j k} = i_{j+1 k} = 0 \hbox{ for any }
1 \le j \le a-1, 1 \le k \le m,
&(4.18{\rm a})\cr
&i_{j k} \prec i_{j k+1} \hbox{ or }
i_{j k}  = i_{j k+1} \in J \setminus \{ 0 \}\,\,
\hbox{ for any } 1 \le j \le a, 1 \le k \le m-1.
&(4.18{\rm b})\cr}
$$
Notice that (4.18a) is equivalent to saying that
each column belongs to ${\cal T}^{(a)}_1$
defined in (4.7).
We identify each element of ${\cal T}^{(a)}_m$ as above
with the following product of (4.4a):
$$
\prod_{j=1}^a\prod_{k=1}^m \Fsquare(0.5cm,i_{j k})
\vert_{u \rightarrow u+a-m-2j+2k}.\eqno(4.19)
$$
Then we conjecture that the $T$-system (2.5a) with 
(4.17) and the initial condition (4.9) leads to
$$
\Lambda^{(a)}_m(u) = 
{1\over \prod_{k=1}^m F^{(p,r)}_a(u-m-1+2k)}
\sum_{T \in {\cal T}^{(a)}_m} T\qquad
1 \le a \le r-1, m \in {\bf Z}_{\ge 1}.
\eqno(4.20)
$$
From (4.18a) and the remark after (4.10),
the rhs is homogeneous of degree $2pm$ w.r.t $\phi$.
The conjecture (4.20) reduces to (4.9a) when $m=1$.
For $B_2$, (4.20) is certainly true 
because $\Lambda^{(1)}_m(u)$ of $B_2$ equals
$\Lambda^{(2)}_m(u)$ of $C_2$ given in (3.27)
under the exchange $Q_1(u) \leftrightarrow Q_2(u)$.
The cases $m=2, a=1,2$ have also been checked directly for 
$B_3$ and $B_4$.
As a further support, we have verified 
$\sharp {\cal T}^{(a)}_m = \hbox{dim } W^{(a)}_m$
by computer for several values of $a$ and $m$.
For example, both sides yields 247500 for $B_5, a=m=3$.
We emphasize that the set ${\cal T}^{(a)}_m$ is
specified by a remarkably simple rule (4.18).
It would deserve to be a proper base of the $U_q(B^{(1)}_r)$ or
$Y(B_r)$ module $W^{(a)}_m$ having the classical content (2.4b).
\par\noindent
{\bf 4.5 Eigenvalue $\Lambda^{(r)}_1(u)$.}\hskip0.3cm
From (2.4b) 
the relevant auxiliary space is $W^{(r)}_1 \simeq V(\omega_r)$
as a $B_r$-module.
This is the spin representation, whose weights are 
all multiplicity-free and given by
$$
{1 \over 2}(\mu_1 \epsilon_1 + \cdots + \mu_r \epsilon_r),
\qquad \mu_1, \ldots, \mu_r = \pm.\eqno(4.21)
$$
Thus we shall introduce another kind of
elementary boxes 
$\Flect(0.4cm,2.5cm,{\mu_1,\mu_2,\ldots,\mu_r})$
by which the DVF can be written as
$$
\Lambda^{(r)}_1(u)=\sum_{\{\mu_j = \pm\}} 
\overbrace{\Flect(0.4cm,2.5cm,{\mu_1,\mu_2, \cdots , \mu_r})}^{r}_p.
\eqno(4.22)
$$
We let ${\cal T}^{(r)}_1$ denote the set of
$\hbox{dim } W^{(r)}_1 = 2^r$ boxes as above.
The indices $r$ and $p$ here signify the rank of $B_r$ and 
the quantum space 
$\otimes_{j=1}^N W^{(p)}_1(w_j)$, respectively.
Each box is identified with a product of dress and vacuum parts
that are defined via certain 
recursion relations w.r.t these indices.
To describe them we introduce the 
operators $\tau^u_\gamma, \tau^Q$ and 
$\tau^C_\gamma$ acting on the DVF (2.9b) as follows.
$$\eqalignno{
\tau^u_\gamma &: u \rightarrow u + \gamma,&(4.23{\rm a})\cr
\tau^Q &: Q_a(u) \rightarrow Q_{a+1}(u),&(4.23{\rm b})\cr
\tau^C_\gamma &: Q_a(u+x) \rightarrow Q_a(u+\gamma-x), \,
                 \phi(u+x) \rightarrow \phi(u+\gamma-x)
                 \hbox{ for any } x.&(4.23{\rm c})\cr}
$$
By the definition they obey the relations
$$\eqalignno{
&\tau^Q \tau^u_\gamma = \tau^u_\gamma \tau^Q,
\quad
\tau^Q \tau^C_\gamma = \tau^C_\gamma \tau^Q,&(4.24{\rm a})\cr
&\tau^C_\gamma \tau^C_{\gamma^\prime} =
\tau^u_{\gamma - \gamma^\prime},
\quad
\tau^u_{\gamma} \tau^u_{\gamma^\prime} = 
\tau^u_{\gamma + \gamma^\prime}.&(4.24{\rm b})\cr
}$$
In view of (2.8), $\tau^Q$ is equivalent to 
$N_a \rightarrow N_{a+1}$ and $u^{(a)}_j \rightarrow u^{(a+1)}_j$.
It is to be understood as replacing $Q_a(u)$ with 
$1 \le a \le r-1$ for $B_{r-1}$ by $Q_{a+1}(u)$ for $B_r$.
The operator $\tau^C_\gamma$ will be used to describe the
transformation (2.19) concerning the crossing symmetry.
Now the recursion relations read,
$$\eqalignno{
\overbrace{\Flect(0.4cm, 1.5cm, {+,+,\xi })}^{r}_p &=
\phi(u+r+p-{3\over 2}+{1\over t_p})
\tau^Q \overbrace{\Flect(0.4cm, 1.2cm, {+,\xi })}^{r-1}_{p-1},
&(4.25{\rm a})\cr
\overbrace{\Flect(0.4cm, 1.5cm, {+,-,\xi })}^{r}_p &=
\phi(u+r+p-{3\over 2}+{1\over t_p})
{Q_1(u+r-{5\over 2})\over Q_1(u+r-{1\over 2})}
\tau^Q \overbrace{\Flect(0.4cm, 1.2cm, {-,\xi })}^{r-1}_{p-1},
&(4.25{\rm b})\cr
\overbrace{\Flect(0.4cm, 1.5cm, {-,+,\xi })}^{r}_p &=
\phi(u+r-p+{1\over 2}-{1\over t_p})
{Q_1(u+r+{3\over 2})\over Q_1(u+r-{1\over 2})}
\tau^u_2 \tau^Q \overbrace{\Flect(0.4cm, 1.2cm, {+,\xi })}^{r-1}_{p-1},
&(4.25{\rm c})\cr
\overbrace{\Flect(0.4cm, 1.5cm, {-,-,\xi })}^{r}_p &=
\phi(u+r-p+{1\over 2}-{1\over t_p})
\tau^u_2 \tau^Q \overbrace{\Flect(0.4cm, 1.2cm, {-,\xi })}^{r-1}_{p-1},
&(4.25{\rm d})\cr
}$$ 
where $\xi$ denotes arbitrary sequence of $\pm$ symbols with length
$r-2$.
The recursions (4.25) are valid for 
$1 \le p \le r$ and $r \ge 3$.
The initial condition is given by
$$\eqalign{
dr \overbrace{\Flect(0.4cm,0.8cm,{+,+})}^{2}_{p} 
   &={Q_2(u-{1\over 2}) \over Q_2(u+{1\over 2})},
   \qquad\qquad  \quad\,
vac \overbrace{\Flect(0.4cm,0.8cm,{+,+})}^{2}_{p} 
= \cases{\phi(u+{5\over 2}) & $p=1$\cr
         \phi(u+1)\phi(u+3) & $p=2$\cr},\cr
dr \overbrace{\Flect(0.4cm,0.8cm,{+,-})}^{2}_{p} 
   &={Q_1(u-{1\over 2})Q_2(u+{3\over 2}) \over 
      Q_1(u+{3\over 2})Q_2(u+{1\over 2})},
   \quad
vac \overbrace{\Flect(0.4cm,0.8cm,{+,-})}^{2}_{p} 
= \cases{\phi(u+{5\over 2}) & $p=1$\cr
         \phi(u)\phi(u+3) & $p=2$\cr},\cr
dr \overbrace{\Flect(0.4cm,0.8cm,{-,+})}^{2}_{p} 
   &={Q_1(u+{7\over 2})Q_2(u+{3\over 2}) \over 
      Q_1(u+{3\over 2})Q_2(u+{5\over 2})},
   \quad
vac \overbrace{\Flect(0.4cm,0.8cm,{-,+})}^{2}_{p} 
= \cases{\phi(u+{1\over 2}) & $p=1$\cr
         \phi(u)\phi(u+3) & $p=2$\cr},\cr
dr \overbrace{\Flect(0.4cm,0.8cm,{-,-})}^{2}_{p} 
   &={Q_2(u+{7\over 2}) \over Q_2(u+{5\over 2})},
   \qquad\qquad  \quad\,
vac \overbrace{\Flect(0.4cm,0.8cm,{-,-})}^{2}_{p} 
= \cases{\phi(u+{1\over 2}) & $p=1$\cr
         \phi(u)\phi(u+2) & $p=2$\cr}.\cr}
\eqno(4.26{\rm a})
$$
Note that one formally needs the dress and the vacuum parts 
for $p=0$ when applying (4.25) with $p=1$.
As for the vacuum parts we fix this by putting
$$
vac \overbrace{\Flect(0.4cm,2.0cm,{\mu_1,\ldots,\mu_r})}^r_0 = 1
\quad \hbox{ for any } r \hbox{ and }  \{ \mu_j \}.
\eqno(4.26{\rm b})
$$
As for the dress parts  we simply let 
$dr \overbrace{\Flect(0.4cm,2.0cm,{\mu_1,\ldots,\mu_r})}^r_p$
be the same for any $0 \le p \le r$.
This is consistent with (4.26a) and the dress universality (2.11).
Under these setting 
the recursions (4.25) and the initial condition (4.26) provide
a complete characterization of our
$\overbrace{\Flect(0.4cm,2.0cm,{\mu_1,\ldots,\mu_r})}^r_p$
for any $0 \le p \le r, r \ge 2$ and 
$\{ \mu_j \}$.
Thus we have presented the DVF (4.22) for the eigenvalue
$\Lambda^{(r)}_1(u)$.
In the rational case ($q \rightarrow 1$) 
with $p=1$, a similar recursive
description is available in [5].
\par
Let us observe various features of our DVF (4.22) before
proving that it is pole-free in section 4.6.
Firstly, it is easy to calculate the vacuum parts
explicitly.
$$\eqalignno{
&vac \overbrace{\Flect(0.4cm,2.0cm,{\mu_1,\ldots,\mu_r})}^r_p
= \psi^{(p,r)}_n(u),&(4.27{\rm a})\cr
&n = \sharp\{ j \mid \mu_j = -, 1 \le j \le p \},&(4.27{\rm b})\cr
&\psi^{(p,r)}_n(u) = 
\prod_{j=0}^{n-1}\phi(u+r-p+2j+{1\over 2} - {1\over t_p})\cr
&\quad\qquad 
\times \prod_{j=n}^{p-1}\phi(u+r-p+2j+{1\over 2} + {1\over t_p})\quad
0 \le n \le p.&(4.27{\rm c})\cr}
$$
This is order $p$ w.r.t $\phi$.
Secondly, 
the top term 
is given by 
$$
\overbrace{\Flect(0.4cm,2.4cm,{+,+,\ldots, +})}^r_p = 
\psi^{(p,r)}_0(u){Q_r(u-{1\over 2})\over Q_r(u+{1\over 2})}.
\eqno(4.28)
$$
This is consistent with (2.12) since
the above box is associated with the highest weight
$(\epsilon_1 + \cdots + \epsilon_r)/2 = \omega_r$ from (4.1) and (4.21).
Thirdly, the crossing symmetry 
$\Lambda^{(r)}_1(u) = 
(-)^{pN}\Lambda^{(r)}_1(-2r+1-u)
\vert_{w_j \rightarrow -w_j, u^{(a)}_i \rightarrow -u^{(a)}_i}$
is valid, which is precisely (2.17) with
the order $k=p$ as remarked above.
At the level of the boxes, this is due to
$$
\tau^C_{2r-1} 
\overbrace{\Flect(0.4cm,2.0cm,{\mu_1,\ldots,\mu_r})}^r_p
= \overbrace{\Flect(0.4cm,2.3cm,{-\mu_1,\ldots,-\mu_r})}^r_p,
\eqno(4.29)
$$
where the effect of $(-)^{pN}$ has been absorbed into 
$\tau^C_{2r-1}$ as explained in (2.19).
In the sequel, we will write such $\pm$ sequences as above
simply as $\mu$ and $\overline{\mu}$, etc. 
As an warming-up exercise 
let us show (4.29) by induction on $r$.
In view of $(\tau^C_\gamma)^2 = 1$, it suffices to check
the two cases $(\mu_1, \mu_2) = (+,+)$ and $(+,-)$.
We shall do the former case 
and leave the latter to the readers.
Put $\mu = (+,+,\nu)$ with $\nu$ being a length $r-2$ sequence of 
$\pm$.
Then the lhs of (4.29) becomes
$$\eqalign{
\tau^C_{2r-1} \overbrace{\Flect(0.4cm,2.0cm,{+,+,\nu})}^r_p
&= \tau^C_{2r-1}\Bigl(
\phi(u+r+p-{3\over 2} +{1\over t_p})\tau^Q
\overbrace{\Flect(0.4cm,1.8cm,{+,\nu})}^{r-1}_{p-1}\Bigr)\cr
&= \phi(u+r-p+{1\over 2}-{1\over t_p})\tau^C_{2r-1}\tau^Q
\tau^C_{2r-3}
\overbrace{\Flect(0.4cm,1.8cm,{-,\overline{\nu}})}^{r-1}_{p-1},\cr
}
$$
where we have used (4.25a) in the first line and 
the induction assumption in the second line.
By means of (4.24) one may substitute
$\tau^C_{2r-1} \tau^Q \tau^C_{2r-3} = \tau^u_2 \tau^Q$
into the latter.
From (4.25d), the result is equal to
$\overbrace{\Flect(0.4cm,2.0cm,{-,-,\overline{\nu}})}^r_p$,
which is the rhs of (4.29).
\par\noindent
{\bf 4.6 Pole-freeness of $\Lambda^{(r)}_1(u)$.}\hskip0.3cm
In section 4.5, we have formally allowed $p=0$ in 
the boxes that consist of the DVF (4.22).
Correspondingly, we find it convenient to consider
the BAE with $p=0$ as the one obtained from 
(2.7) by setting its lhs always $-1$.
We shall quote (2.7) as $\hbox{BAE}^r_p$.
Our aim here is to establish
\proclaim Theorem 4.6.1.
For $r \ge 2$ and $0 \le p \le r$, $\Lambda^{(r)}_1(u)$ (4.22)
is free of poles provided that the $\hbox{BAE}^r_p$ (2.7) 
(for $s=1$) is valid.
\par
We are to show that color $a$ poles 
$1/Q_a$ are spurious for each $1 \le a \le r$.
The poles are located by 
\proclaim Lemma 4.6.2.
For $1 \le a \le r-1$ the factor $1/Q_a$
is contained in the box
$\overbrace{\Flect(0.4cm,2.0cm,{\mu_1,\ldots,\mu_r})}^r_p$
if and only if $(\mu_a,\mu_{a+1}) = (+,-)$ or $(-,+)$.  
Any two such boxes
$\overbrace{\Flect(0.4cm,2.0cm,{\eta,+,-,\xi})}^r_p$ and 
$\overbrace{\Flect(0.4cm,2.0cm,{\eta,-,+,\xi})}^r_p$
share a common color $a$ pole $1/Q_a(u+y)$ for some $y$.
The factor $1/Q_r$ is contained in all the boxes.
For $\epsilon = \pm$, any two boxes
$\overbrace{\Flect(0.4cm,1.5cm,{\zeta,\epsilon,\epsilon})}^r_p$ and 
$\overbrace{\Flect(0.4cm,1.5cm,{\zeta,\epsilon,-\epsilon})}^r_p$ 
share a common color $r$ pole $1/Q_r(u+z)$ for some $z$.
\par
The assertions are immediate consequences of (4.25) and (4.26).
If one puts 
\par\noindent
$\lambda = (\eta,+,-,\xi)$, 
$\mu = (\eta,-,+,\xi)$ and identifies them with the weights via (4.21), 
one has $\lambda - \mu = \epsilon_a - \epsilon_{a+1} = \alpha_a$
for $1 \le a \le r-1$ by (4.21).
A similar relation holds for $a=r$ as well.
Thus the above Lemma is another example of the 
coupling rule (2.14a).
In this view 
Theorem 4.6.1 is a corollary of 
\proclaim Theorem 4.6.3.
For $1 \le a \le r-1$, 
let $\eta, \xi$ and $\zeta$ be any $\pm$ sequences with
lengths $a-1, r-a-1$ and $r-2$, respectively.
If the $\hbox{BAE}^r_p$ (2.7) (for $s=1$) is valid, then
$$\eqalignno{
&Res_{u=-y+iu^{(a)}_k}\Bigl(
\overbrace{\Flect(0.4cm,2.0cm,{\eta,+,-,\xi})}^r_p +
\overbrace{\Flect(0.4cm,2.0cm,{\eta,-,+,\xi})}^r_p \Bigr) = 0,
&(4.30{\rm a})\cr
&Res_{u=-z+iu^{(r)}_k}\Bigl(
\overbrace{\Flect(0.4cm,2.0cm,{\zeta,\pm,\pm})}^r_p +
\overbrace{\Flect(0.4cm,2.0cm,{\zeta,\pm,\mp})}^r_p \Bigr) = 0,
&(4.30{\rm b})\cr
}$$
where $y$ and $z$ are those in Lemma 4.6.2.
\par
The rest of the present subsection will be devoted to a proof
of this Theorem assuring that all the color $a$ poles are spurious.
In fact the proof will be done essentially by establishing 
(2.14b) and (2.15).
It follows that the character limit (2.16) is also valid for 
$\Lambda^{(r)}_1(u)$ (4.22).
We prepare 
\proclaim Lemma 4.6.4.
Let $\xi$ be any sequence of $\pm$ with length $r-1$.
Then
$$\tau^C_{2r+1} 
\overbrace{\Flect(0.5cm,1.3cm,{-,\overline{\xi}})}^r_p
= \Bigl({\phi(u+r+p+{1\over 2}+{1\over t_p})\over
         \phi(u+r-p+{1\over 2}-{1\over t_p})}\Bigr)^{1-\delta_{p 0}}
{Q_1(u+r-{1\over 2})\over Q_1(u+r+{3\over 2})}
\overbrace{\Flect(0.5cm,1.3cm,{-,\xi})}^r_p
\eqno(4.31)
$$
for $0 \le p \le r$.
\par\noindent
{\it Proof.}\hskip0.3cm
We show this for $\xi = (+,\nu)$.
The case $\xi = (-,\nu)$ is similar.
Suppose $p \ge 1$. Then the lhs of (4.31) is rewritten as
$$\eqalign{
\tau^C_{2r+1} \overbrace{\Flect(0.4cm,1.5cm,{-,-,\overline{\nu}})}^r_p
&= \tau^C_{2r+1}\tau^C_{2r-1}
\overbrace{\Flect(0.4cm,1.5cm,{+,+,\nu})}^r_p\cr
&= \phi(u+r+p+{1\over 2}+{1\over t_p})\tau^u_2\tau^Q
\overbrace{\Flect(0.4cm,1.5cm,{+,\nu})}^{r-1}_{p-1}\cr}
\eqno(4.32)
$$
by means of (4.24b), (4.25a) and (4.29).
In the rhs of (4.31), the box part is 
$\overbrace{\Flect(0.4cm,1.5cm,{-,+,\nu})}^r_p$.
Replacing this by the rhs of (4.25c) with 
$\xi = \nu$, one finds that the resulting
expression coincides with the last line in (4.32).
The case $p=0$ follows from this and (4.26b).
\par
Finally we give
\par\noindent
{\it Proof of Theorem 4.6.3.}\hskip0.3cm
It is straightforward to check (4.30) for $r=2$ 
by (4.26a).
We assume that Theorem is true for $B_{r-1}$ 
and use induction on $r$.
We shall verify (4.30a) only.
Eq. (4.30b) can be shown more easily by a similar method.
In the sequel the cases $a \ge 3$, $a=2$ and $a=1$
are considered separately.
\par
{\it The case $a \ge 3$.}\hskip0.3cm
Put $\eta = (\eta_1,\eta^\prime)$.
Then (4.25) transforms the sum of the two boxes 
in (4.30a) into
$$
X_1 \tau^u_{1-\eta_1} \tau^Q\Bigl(
\overbrace{\Flect(0.4cm,2.0cm,{\eta^\prime,+,-,\xi})}^{r-1}_{p-1}
+ \overbrace{\Flect(0.4cm,2.0cm,{\eta^\prime,-,+,\xi})}^{r-1}_{p-1}
\Bigr),
$$
where $X_1$ involves only $\phi$ and possibly $Q_1/Q_1$.
Then Lemma 4.6.2 implies that
the poles $1/Q_a(u+y)$ in (4.30a) must originate in the factor 
$1/Q_{a-1}(u+y-1+\eta_1)$ shared by the above two boxes.
Thus the lhs of (4.30a) is proportional
to $Res_{u=-y+1-\eta_1 + iu^{(a-1)}_k}$ of the sum of the above
boxes.
But this is a case of (4.30a) for $B_{r-1}$ hence $0$ by
the induction assumption.
\par
{\it The case $a = 2$.}\hskip0.3cm
In (4.30a) the length of $\eta$ is $1$. 
We consider the case $\eta = +$.
The proof for $\eta=-$ is almost identical.
First we rewrite the two boxes in (4.30a) by (4.25) as follows.
$$
\overbrace{\Flect(0.4cm,1.8cm,{+,+,-,\xi})}^r_p = X_2
\tau^Q
\overbrace{\Flect(0.4cm,1.5cm,{+,-,\xi})}^{r-1}_{p-1},\quad
\overbrace{\Flect(0.4cm,1.8cm,{+,-,+,\xi})}^r_p
= X_2
{Q_1(u+r-{5\over 2})\over Q_1(u+r-{1\over 2})}
\tau^Q \overbrace{\Flect(0.4cm,1.5cm,{-,+,\xi})}^{r-1}_{p-1},
\eqno(4.33)
$$
where $X_2 = \phi(u+r+p-{3\over 2}+{1\over t_p})$
is independent of $Q_b$'s.
By further using (4.25), one finds that 
$$
\overbrace{\Flect(0.4cm,1.5cm,{+,-,\xi})}^{r-1}_{p-1} 
= X_3 {Q_1(u+r-{7\over 2})\over Q_1(u+r-{3\over 2})},\quad
\overbrace{\Flect(0.4cm,1.5cm,{-,+,\xi})}^{r-1}_{p-1}
= X_4 {Q_1(u+r+{1\over 2})\over Q_1(u+r-{3\over 2})},
\eqno(4.34)
$$
where $X_3$ and $X_4$ do not involve $Q_1$.
From the induction assumption,
the sum of these boxes must be ensured to be regular
at $u = -r+{3\over 2} +iu^{(1)}_k$ 
via the $\hbox{BAE}^{r-1}_{p-1}$.
Therefore
$$\eqalign{
&\overbrace{\Flect(0.4cm,1.5cm,{-,+,\xi})}^{r-1}_{p-1} /
\overbrace{\Flect(0.4cm,1.5cm,{+,-,\xi})}^{r-1}_{p-1}
\cr
&\quad = 
{Q_1(u+r+{1\over 2})Q_2(u+r-{5\over 2})
\phi(u+r-{3\over 2} - {1\over t_p}\delta_{1 p-1}) \over
Q_1(u+r-{7\over 2})Q_2(u+r-{1\over 2})
\phi(u+r-{3\over 2} + {1\over t_p}\delta_{1 p-1})},\cr}
\eqno(4.35)
$$
should hold in order that the 
lhs be evaluated as $-1$ at $u = -r + {3\over 2} + iu^{(1)}_k$ 
from the $\hbox{BAE}^{r-1}_{p-1}$.
In deriving (4.35)
we have used the fact that $t_{p-1}$ of $B_{r-1}$ is equal to
$t_p$ of $B_r$.
Combining (4.33) and (4.35) one deduces
$$\eqalign{
&\overbrace{\Flect(0.4cm,1.8cm,{+,-,+,\xi})}^{r}_{p} /
\overbrace{\Flect(0.4cm,1.8cm,{+,+,-,\xi})}^{r}_{p}
\cr
&\quad = 
{Q_1(u+r-{5\over 2})Q_2(u+r+{1\over 2})Q_3(u+r-{5\over 2})
\phi(u+r-{3\over 2} - {1\over t_p}\delta_{2 p}) \over
Q_1(u+r-{1\over 2})Q_2(u+r-{7\over 2})Q_3(u+r-{1\over 2})
\phi(u+r-{3\over 2} + {1\over t_p}\delta_{2 p})}.\cr}
\eqno(4.36)
$$
Thanks to $\hbox{BAE}^r_p$ (2.7), this is indeed $-1$
at $u = -r + {3\over 2} + iu^{(2)}_k$,
proving that color $a=2$ poles are spurious.
\par
{\it The case $a=1$.}\hskip0.3cm
The two boxes in (4.30a) are 
$\overbrace{\Flect(0.4cm,1.5cm,{+,-,\xi})}^{r}_{p}$ and 
$\overbrace{\Flect(0.4cm,1.5cm,{-,+,\xi})}^{r}_{p}$.
From (4.25b,c) they share a color $a=1$ pole
at $u=-r+{1\over 2}+iu^{(1)}_k$.
Let us rewrite the latter as follows.
$$\eqalign{
\overbrace{\Flect(0.5cm,1.5cm,{-,+,\xi})}^{r}_{p}
&= \tau^C_{2r-1}
\overbrace{\Flect(0.5cm,1.5cm,{+,-,\overline{\xi}})}^{r}_{p}\cr
&= \phi(u+r-p+{1\over 2}-{1\over t_p})
{Q_1(u+r+{3\over 2}) \over Q_1(u+r-{1\over 2})}
\tau^Q\tau^C_{2r-1}
\overbrace{\Flect(0.5cm,1.2cm,{-,\overline{\xi}})}^{r-1}_{p-1},\cr}
$$
where we have used (4.29) and (4.25b).
In the last line, 
$\tau^C_{2r-1}
\overbrace{\Flect(0.5cm,1.2cm,{-,\overline{\xi}})}^{r-1}_{p-1}$
can be further rewritten by applying Lemma 4.6.4 with 
$r \rightarrow r-1, p \rightarrow p-1$.
Dividing the resulting expression by the rhs of 
(4.25b) we obtain
$$\eqalign{
\overbrace{\Flect(0.4cm,1.5cm,{-,+,\xi})}^{r}_{p} /
\overbrace{\Flect(0.4cm,1.5cm,{+,-,\xi})}^{r}_{p}
= 
{Q_1(u+r+{3\over 2})Q_2(u+r-{3\over 2})\over
 Q_1(u+r-{5\over 2})Q_2(u+r+{1\over2})}
\Bigl(
{\phi(u+r-p+{1\over 2}-{1\over t_p})\over
 \phi(u+r+p-{3\over 2}+{1\over t_p})}
\Bigr)^{\delta_{1 p}}.}
$$
At the pole location $u=-r+{1\over 2}+iu^{(1)}_k$,
this is just $-1$ owing to $\hbox{BAE}^r_p$ (2.7) 
with $a=1$.
Therefore (4.30a) is free of color $1$ poles.
This completes the proof of Theorem 4.6.3 hence Theorem 4.6.1.
\par\noindent
{\bf 4.7 Relations between two kinds of boxes.}\hskip0.3cm
Here we clarify the relation between the two kinds
of the boxes 
$\Fsquare(0.4cm,a)$ and $\Flect(0.4cm,1.8cm,{\mu_1,\ldots,\mu_r})$
introduced in section 4.1 and 4.5, respectively.
In terms of the relevant auxiliary spaces,
they are associated with the vector 
and the spin representations.
To infer their relation, recall the classical tensor
product decomposition
$$
V(\omega_r)\otimes V(\omega_r)
= V(2\omega_r) \oplus V(\omega_{r-1}) \oplus \cdots \oplus
V(\omega_1) \oplus V(0).
\eqno(4.37)
$$
Correspondingly, there exists an
$U_q(B^{(1)}_r)$ quantum $R$-matrix 
$R_{W^{(r)}_1, W^{(r)}_1}(u)$ [32]
acting on the $q$-analogue of the above.
On each component $V(\omega)$ of the rhs,
it acts as a constant $\rho_\omega(u)$ that depends on 
the spectral parameter $u$.
A little investigation of the spectrum $\rho_\omega(u)$
in [32] tells that only 
$\rho_{\omega_a}(u), \rho_{\omega_{a-2}}(u), \ldots$
are non zero
at $u=-2(r-a)+1$ for $1\le a \le r-1$.
From this and (2.4b) we see that the specialized $R$-matrix 
$R_{W^{(r)}_1, W^{(r)}_1}(-2(r-a)+1)$
yields the embedding
$$
W^{(a)}_1(u) \hookrightarrow W^{(r)}_1(u+r-a-{1\over 2})
\otimes W^{(r)}_1(u-r+a+{1\over 2})\eqno(4.38)
$$
in the notation of [8].
According to the arguments there, (4.38) imposes
the following functional relation among the 
transfer matrices having the relevant auxiliary spaces:
$$
T^{(r)}_1(u+r-a-{1\over 2})T^{(r)}_1(u-r+a+{1\over 2})
= T^{(a)}_1(u) + T^\prime(u) \quad \hbox{ for } 1 \le a \le r-1.
\eqno(4.39)
$$
Here $T^\prime(u)$ denotes some matrix commuting with all
$T^{(b)}_m(v)$'s.
When $a = r-1$, (4.39) is 
just the last equation in (2.5a) with $m=0$, hence
$T^\prime(u) = T^{(r)}_2(u)$.
Viewed as a relation among the eigenvalues, 
(4.39) implies that each term in the DVF (4.9a)
can be expressed as a product of 
certain two boxes in section 4.5 with the spectral parameters
$u+r-a-{1\over 2}$ and $u-r+a+{1\over 2}$.
Actually we have
\proclaim Theorem 4.7.1.
For $1 \le a \le r-1, k, n, l \in {\bf Z}_{\ge 0}$ such that 
$k+n+l=a$, take any integers
$1 \le i_1 < \cdots < i_k \le r$ and 
$1 \le j_1 < \cdots < j_l \le r$.
Then the following equality holds between the elements of
${\cal T}^{(a)}_1$ and ${\cal T}^{(r)}_1$ defined in 
(4.7,8) and (4.25,26), respectively.
$$
\raise 13ex \hbox{${1 \over F^{(p,r)}_a(u)}$}
\,
\hbox{
   \normalbaselines\m@th\baselineskip0pt\offinterlineskip
   \vbox{ 
      \hbox{$\Fsquare(0.5cm,\hbox{$i_1$})$}\vskip-0.4pt
      \hbox{$\vnaka$}\vskip-0.4pt
	     \hbox{$\Fsquare(0.5cm,\hbox{$i_k$})$}\vskip-0.4pt
	     \os
      \hbox{$\Fsquare(0.5cm,\hbox{$\overline{j}_{\ell}$})$}\vskip-0.4pt
      \hbox{$\vnaka$}\vskip-0.4pt
      \hbox{$\Fsquare(0.5cm,\hbox{$\overline{j}_1$})$} 
        }
      }
\,\, 
 \raise 13ex 
  \hbox{$= \bigl(\, \tau^u_{-r+a+{1\over 2}}
        \overbrace{\Flect(0.4cm,1.8cm,{\mu_1,\ldots,\mu_r})}^r_p \, \bigr)
        \bigl(\, \tau^u_{r-a-{1\over 2}}
        \overbrace{\Flect(0.4cm,1.8cm,{\nu_1,\ldots,\nu_r})}^r_p \, \bigr),
    $}\eqno(4.40{\rm a})
$$
where there are $n$ $\Fsquare(0.4cm,0)$'s in the lhs and
$F^{(p,r)}_a(u)$ is defined in (4.9b) and (4.10).
The $\pm$ sequences in the rhs are specified by
$$\eqalign{
\mu_b &= \cases{+ & if $b \in \{ i_1, \ldots, i_k \}$ \cr
                - & otherwise \cr},\cr
\nu_b &= \cases{- & if $b \in \{ j_1, \ldots, j_l \}$ \cr
                + & otherwise \cr}.\cr}\eqno(4.40{\rm b})
$$
\par
Note that both sides of (4.40a) are of order
$2p$ w.r.t $\phi$ and carry the same weight
$\epsilon_{i_1} + \cdots + \epsilon_{i_k} 
- \epsilon_{j_1} - \cdots - \epsilon_{j_l}$.
The Theorem is again proved by induction on the rank $r$.
To do so we write the boxes (4.4a) as
$\overbrace{\Fsquare(0.4cm,a)}^r_p$ to mark the 
$r, p$ dependence explicitly.
Then they enjoy the recursive property as follows.
\proclaim Lemma 4.7.2.
For $1 \le p \le r$ and $2 \le b \le r$, the boxes (4.4) 
fulfill 
$$\eqalign{
\overbrace{\Fsquare(0.4cm,b)}^r_p &= X_1 
\Bigl({Q_1(u+3)\over Q_1(u+1)}\Bigr)^{\delta_{2 b}} 
\tau^u_1 \tau^Q
\overbrace{\Flect(0.4cm,1.0cm,b-1)}^{r-1}_{p-1},\cr
X_1 &= \phi(u-p+2-{1\over t_p})\phi(u+2r+p-3+{1\over t_p}).\cr}
\eqno(4.41)
$$
\par
This can be checked directly by using the explicit form
(4.4).
In particular, one uses
$X_1 = \Phi^r_p(u)/\Phi^{r-1}_{p-1}(u+1)$.
\par\noindent
{\it Proof of Theorem 4.7.1 for $a=k=1$ and $n=l=0$.}
We illustrate an inductive proof w.r.t $r$ in this case.
General cases can be verified based on it by a similar idea
through tedious calculations.
Eq.(4.40) can be directly checked for $r=2$.
By letting $i_1 = b$ and noting that 
$F^{(p,r)}_1(u) = 1$, (4.40) reads
$$
\overbrace{\Fsquare(0.4cm,b)}^r_p = 
\bigl(\, \tau^u_{-r+{3\over 2}}
\overbrace{\Flect(0.4cm,3.7cm,{-,\cdots,-,+,-,\cdots,-})}^r_p \,\bigr)
\bigl(\, \tau^u_{r-{3\over 2}}
\overbrace{\Flect(0.4cm,1.5cm,{+,\cdots,+})}^r_p \,\bigr),
\eqno(4.42)
$$
where the $+$ symbol in the first box on the rhs is located 
at $b$-th position from the left.
Below we focus on the case $b \ge 2$
and leave $b=1$ case to the readers.
Then, by applying the recursions (4.25) and 
(4.24) the rhs is rewritten as
$$
X_1 \Bigl({Q_1(u+3) \over Q_1(u+1)} \Bigr)^{\delta_{2 b}}
\tau^u_1\tau^Q\Biggl(
\bigl(\, \tau^u_{-r+{5\over 2}}
\overbrace{
\Flect(0.4cm,3.7cm,{-,\cdots,-,+,-,\cdots,-})}^{r-1}_{p-1}\,\bigr)
\bigl(\, \tau^u_{r-{5\over 2}}
\overbrace{\Flect(0.4cm,1.5cm,{+,\cdots,+})}^{r-1}_{p-1} \,\bigr)
\Biggr),\eqno(4.43)
$$
where $X_1$ is the one in (4.41).
The $+$ symbol in the first box is now at the $b-1$ th position.
The quantity in the largest parenthesis of (4.43) is precisely
the rhs of (4.42) with $r \rightarrow r-1$ and $p \rightarrow p-1$.
By induction one may replace it with 
$\overbrace{\Flect(0.4cm,1.0cm,b-1)}^{r-1}_{p-1}$.
The resulting expression is just the rhs of (4.41) hence
equal to $\overbrace{\Fsquare(0.4cm,b)}^r_p$ by Lemma 4.7.2.
This completes the induction step hence the proof.
\beginsection 5. Eigenvalues for $D_r$

Our results for $D_r = so(2r)$ are quite parallel with those
for $B_r = so(2r+1)$ in many respects.
In fact many formulas here
becomes those in section 4 through a formal 
replacement $r \rightarrow r + {1\over 2}$.
Thus we shall state them without a proof,
which can be done in a similar manner to the $B_r$ case.
We will introduce two kinds of boxes 
associated with the vector and the spin representations
and clarify their relation.
A distinct feature in $D_r$ is 
that there are two representations of the latter kind,
$V(\omega_{r-1})$ and $V(\omega_r)$, each having 
the quantum affine analogue 
$W^{(r-1)}_1$ and $W^{(r)}_1$, respectively.
They are interchanged under the Dynkin diagram automorphism.
In order to respect the symmetry under it,
we modify the 
quantum spaces 
$\otimes_{j=1}^N W^{(p)}_1(w_j)$ 
for $p=r-1$ and $r$ into
$\otimes_{j=1}^N W^{(\pm)}_1(w_j)$ where
$W^{(\pm)}_1(w) = W^{(r)}_1(w\mp 2)\otimes W^{(r-1)}_1(w\pm 2)$.
Pictorially, one may view this as
arranging the vertical lines on the square lattice 
endowed with the modules 
$V(\omega_r), V(\omega_{r-1})$ alternately and with 
the inhomogeneity as
$w_1 \mp 2, w_1 \pm 2, w_2 \mp 2, w_2 \pm 2, \ldots$. 
This pattern has been introduced to utilize the degeneracy of
the spin-conjugate spin $R$-matrix [32]
$\hbox{Im} R_{W^{(r)}_1, W^{(r-1)}_1}(u=4) \simeq
V(\omega_r + \omega_{r-1})$,
where the image becomes manifestly symmetric under the automorphism.
The BAE (2.7) (with $s=1$) is thereby 
unchanged as long as $p = 1,2, \ldots, r-2$.
Instead of $p=r-1$ and $r$, we now take
$p=\pm$, for which the BAE reads
$$
-{\phi^+_p(iu_k^{(a)}+\delta_{a r}) \phi^-_p(iu_k^{(a)}+\delta_{a r-1})
\over
 \phi^+_p(iu_k^{(a)}-\delta_{a r})  \phi^-_p(iu_k^{(a)}-\delta_{a r-1}) } =
\prod_{b=1}^r {{Q_b(iu_k^{(a)}+(\alpha_a|\alpha_b)) }\over
                {Q_b(iu_k^{(a)}-(\alpha_a|\alpha_b))}}.\eqno(5.1)
$$
Here the functions in the lhs are defined via $\phi(u)$ (1.4b) by
$$\phi^{\pm}_{\pm}(u) = \phi(u+2), \quad
  \phi^{\pm}_{\mp}(u) = \phi(u-2).\eqno(5.2)
$$
\par\noindent
{\bf 5.1 Eigenvalue $\Lambda^{(1)}_1(u)$.}\hskip0.3cm
Let $\epsilon_a, 1 \le a \le r$ be the orthonormal vectors 
$(\epsilon_a \vert \epsilon_b) = \delta_{a b}$
realizing the root system as follows.
$$\eqalign{
\alpha_a &= \cases{\epsilon_a - \epsilon_{a+1} &
for $1 \le a \le r-1$\cr
\epsilon_{r-1} + \epsilon_r & for $a = r$\cr},\cr
\omega_a &= 
\cases{
\epsilon_1 + \cdots + \epsilon_a& for $1 \le a \le r-2$\cr
{1 \over 2}(\epsilon_1 + \cdots + \epsilon_{r-1} - \epsilon_r)
& for $a = r-1$\cr
{1 \over 2}(\epsilon_1 + \cdots + \epsilon_{r-1} + \epsilon_r)
& for $a = r$\cr}.\cr}
\eqno(5.3)$$
The auxiliary space relevant to 
$\Lambda^{(1)}_1(u)$ is 
$W^{(1)}_1 \simeq V(\omega_1)$ as an $D_r$-module by (2.4b).
This is the vector representation, whose weights are 
all multiplicity-free and given by
$\epsilon_a$ and $-\epsilon_a (1 \le a \le r)$.
By abbreviating them to $a$ and $\overline{a}$,
the set of weights and the DVF are given as follows.
$$\eqalignno{
J &= \{1,2,\ldots,r,\overline{r},\ldots,\overline{1}\},
&(5.4)\cr
\Lambda^{(1)}_1(u) &= \sum_{a \in J}\Fsquare(0.4cm,a).
&(5.5)\cr}
$$
This is formally the same with (3.2-3).
The elementary boxes are defined by
$$\eqalign{
\Fsquare(0.5cm,a)  &= \psi_{a}(u) 
      {{Q_{a-1}( u+a+1  ) 
            Q_{a}(u+a-2 )}\over
       { Q_{a-1}(u+a-1)Q_{a}(u+a)}} 
  \qquad 1 \le a \le r-2,\cr
\Flect(0.5cm,1.0cm,r-1)  &= \psi_{r-1}(u) 
      {{Q_{r-2}(u+r) Q_{r-1}(u+r-3 ) Q_{r}(u+r-3 )}\over
       { Q_{r-2}(u+r-2) Q_{r-1}(u+r-1 ) Q_{r}(u+r-1 )}}, \cr
\Fsquare(0.5cm, r)  &= 
  \psi_r(u) {{Q_{r-1}(u+r+1) Q_{r}(u+r-3)}\over
   { Q_{r-1}(u+r-1)Q_{r}(u+r-1)}},  \cr
\Fsquare(0.5cm,\overline{r})  &= \psi_{\overline{r}}(u) 
  {{Q_{r-1}(u+r-3) Q_{r}(u+r+1)}\over
   { Q_{r-1}(u+r-1)Q_{r}(u+r-1)}},  \cr
\Flect(0.5cm,1.0cm,\overline{r-1})  &= 
      \psi_{\overline{r-1}}(u) 
      {{Q_{r-2}(u+r-2) Q_{r-1}(u+r+1 ) Q_{r}(u+r+1 )}\over
       { Q_{r-2}(u+r) Q_{r-1}(u+r-1 ) Q_{r}(u+r-1 )}}, \cr
\Fsquare(0.5cm,\overline{a})  &= 
\psi_{\overline{a}}(u) 
      {{Q_{a-1}( u+2r-a-3  ) 
            Q_{a}(u+2r-a)}\over
       { Q_{a-1}(u+2r-a-1)Q_{a}(u+2r-a-2)}} 
  \qquad 1\le a \le r-2,\cr}\eqno(5.6{\rm a})
$$
where we have set $Q_0(u) = 1$.
The vacuum part $\psi_{a}(u)$ depends on the quantum space
$\otimes_{j=1}^N W^{(p)}_1(w_j)$ and is given by 
$$\eqalign{
&\hbox{if }  1 \le p \le r-2 \cr
&\psi_a(u) =
\cases{ \phi(u+p+1) \phi(u+2r-p-1) \Phi^r_p(u)&   
for $1 \preceq a \preceq p$\cr
        \phi(u+p-1) \phi(u+2r-p-1) \Phi^r_p(u)&   
for $p+1 \preceq a \preceq \overline{p+1}$   \cr
        \phi(u+p-1) \phi(u+2r-p-3)\Phi^r_p(u)&
for $\overline{p} \preceq a \preceq \overline{1}$ \cr
                     }  \cr
&\hbox{if }  p =\pm \cr
&\psi_a(u) 
=\cases{\phi^{+}_p(u+r) \phi^{-}_p(u+r) \Phi^r_{r+1}(u)& 
                          for  $1 \preceq a \preceq r-1$ \cr
                   \phi^{+}_p(u+r) \phi^{-}_p(u+r-2) \Phi^r_{r+1}(u)& 
                      for $a=r$ \cr
                   \phi^{-}_p(u+r) \phi^{+}_p(u+r-2) \Phi^r_{r+1}(u)& 
                      for $a=\overline{r}$ \cr
                   \phi^{-}_p(u+r-2) \phi^{+}_p(u+r-2) \Phi^r_{r+1}(u)& 
               for  $\overline{r-1} \preceq a \preceq \overline{1} $ \cr
           }   \cr}\eqno(5.6{\rm b})
$$
where
$$\eqalign{
\Phi^r_p(u) &= \prod_{j=1}^{p-1}
\phi(u+p-2j-1)\phi(u+2r-p+2j-1)\cr
&= \Phi^r_p(-2r+2-u)\vert_{w_k \rightarrow -w_k}.\cr}
\eqno(5.6{\rm c})
$$
The order $\prec$ in the set $J$ (5.4) is specified by
$$
1 \prec 2 \prec \cdots \prec r-1 \prec
\eqalign{&r\cr &\overline{r}\cr} \prec \overline{r-1} \prec 
\cdots \overline{2} \prec \overline{1}.\eqno(5.7)
$$
We impose no order between $r$ and $\overline{r}$.
The DVF (5.5) possesses all the features
explained in section 2.4.
In particular it is pole-free under the BAE (2.7) and (5.1) thanks to
the coupling rule (2.14).
It can be summarized in the diagram
\vskip 0.5cm
\centerline{$\Fsquare(0.4cm,r)$}\par\noindent
\centerline{${\buildrel r-1 \over \nearrow} \hskip0.4cm
{\buildrel r \over \searrow}   $} \par\noindent
\centerline{$
\Fsquare(0.4cm,1) {\buildrel 1 \over \longrightarrow} \Fsquare(0.4cm,2)
{\buildrel 2 \over \longrightarrow} \cdots 
{\buildrel r-2 \over \longrightarrow}
\Flect(0.4cm,1.0cm,r-1)
\hskip 1.2cm
\Flect(0.4cm,1.0cm,\overline{r-1})
{\buildrel r-2 \over \longrightarrow}
\cdots {\buildrel 2 \over \longrightarrow}
\Fsquare(0.4cm,\overline{2})
{\buildrel 1 \over \longrightarrow}
\Fsquare(0.4cm,\overline{1}) 
$}\par\noindent
\centerline{${\buildrel r \over \searrow} \hskip0.4cm
{\buildrel r-1 \over \nearrow}   $} \par\noindent
\centerline{$\Fsquare(0.4cm,\overline{r})$}\par\noindent
\vskip0.5cm\par\noindent
in the same sense with those in sections 3.1 and 4.1.
This is again identical with the crystal graph [24,25].
For $p=1$, the DVF (5.5-6) has been known earlier in [21].
\par\noindent\
{\bf 5.2 Eigenvalue $\Lambda^{(a)}_1(u)$ for $1 \le a \le r-2$.}
\hskip0.3cm
For $1 \le a \le r-2$, let ${\cal T}^{(a)}_1$ be the set of
the tableaux of the form (3.7a) with $i_k \in J$ (5.4) 
obeying the condition
$$
i_k \prec i_{k+1} \hbox{ or } 
(i_k, i_{k+1}) = (r,\overline{r}) \hbox{ or }
(i_k, i_{k+1}) = (\overline{r},r)
\,\,\hbox{ for any } 1 \le k \le a-1.\eqno(5.8)
$$
We identify each element (3.7a) of ${\cal T}^{(a)}_1$ with
the product of (5.6a) with the spectral parameters
$u+a-1, u+a-3,\ldots,u-a+1$ from the top to the bottom as in (4.8).
Then the analytic Bethe ansatz yields the following DVF.
$$
\Lambda^{(a)}_1(u) = 
{1\over F^{(p,r)}_a(u)} 
\sum_{T \in {\cal T}^{(a)}_1} T\quad
\qquad 1 \le a \le r-2.\eqno(5.9{\rm a})
$$
Here the function $F^{(p,r)}_a(u)$ is defined by
$$\eqalign{
&F^{(p,r)}_a(u)\cr
&= \cases{
\prod_{j=1}^{a-1}
\psi^{(p,r)}_0(u+r-a-1+2j)
\psi^{(p,r)}_p(u-r+a+1-2j)& for $1 \le p \le r-2$\cr
\prod_{j=1}^{a-1}\psi^{(p,r)}_{0,+}(u+r-a-1+2j)
\psi^{(p,r)}_{r-1,-}(u-r+a+1-2j)& for $p = \pm$\cr}\cr
&= F^{(p,r)}_a(-2r+2-u)\vert_{w_k \rightarrow -w_k},\cr}
\eqno(5.9{\rm b})
$$
where $\psi^{(p,r)}_n(u)$ and 
$\psi^{(p,r)}_{n,\pm}(u)$ are specified in 
(5.15) in section 5.4.
Notice that $F^{(p,r)}_1(u) = 1$ hence (5.9a)
reduces to (5.5) when $a=1$.
It can be shown that each summand $T$ in (5.9a)
contains the factor $F^{(p,r)}_a$.
This will be seen manifestly in Theorem 5.5.1.
The DVF (5.9) for $\Lambda^{(a)}_1(u)$ is homogeneous w.r.t $\phi$ 
of order $2p$ if $1\le p \le r-2$ and 
order $2r+2$ if $p=\pm$. 
\par
One can observe the top term and the crossing
symmetry in the DVF (5.9) as done after (3.9).
To check the character limit (2.16)
is also similar to (4.11).
From (2.4b) we must show (4.11c) again for $1 \le a \le r-2$ 
under the absence of $y_0$ in (4.11b).
But this is straightforward from (5.8) and by noting that 
the character formula (4.11d) is still valid for $D_r$
if $J$ is taken as (5.4).
\par
By a similar method to Theorem 4.3.1 one can prove
\proclaim Theorem 5.2.1.
$\Lambda^{(a)}_1(u) (1 \le a \le r-2)$ (5.9) is free of poles
provided that the BAE (2.7) (with $s=1$) 
for $1 \le p \le r-2$ and (5.1) for $p=\pm$ are valid.
\par
\par\noindent
{\bf 5.3 Eigenvalue $\Lambda^{(1)}_m(u)$.}
\hskip0.3cm
Starting from (5.9) and the DVFs of 
$\Lambda^{(r)}_1(u), \Lambda^{(r-1)}_1(u)$
that will be given in section 5.4, we are to solve
the $T$-system (2.5c).
The scalar 
$g^{(a)}_m(u)$ there is to be taken as (4.17)
with $F^{(p,r)}_2(u)$ 
determined from (5.9b).
The solution will yield a DVF
for the general eigenvalue $\Lambda^{(a)}_m(u)$.
This program is yet to be executed completely.
Here we shall only present a conjecture on
$\Lambda^{(1)}_m(u)$.
\par
For $m \in {\bf Z}_{\ge 1}$, let ${\cal T}^{(1)}_m$ denote 
the set of tableaux of the form
$$
\Fsquare(0.4cm,i_1)\naga\Fsquare(0.4cm,i_m)
$$
with $i_k \in J$ (5.4) obeying the condition
$$\eqalign{
&i_k \preceq i_{k+1} \quad \hbox{ for any } 1 \le k \le m-1,\cr
&r \hbox{ and } \overline{r} \hbox{ do not appear simultaneously.}\cr
}\eqno(5.10)
$$
We identify each element of ${\cal T}^{(1)}_m$ as above with 
the product of (5.6a) as follows.
$$
\prod_{k=1}^m \Fsquare(0.4cm,i_k)\vert_{u \rightarrow u-m-1+2k}.
$$
Then we have the conjecture
$$
\Lambda^{(1)}_m(u) = \sum_{T \in {\cal T}^{(1)}_m} T,\eqno(5.11)
$$
which reduces to (5.5) when $m=1$.
It is easy to prove
$\sharp {\cal T}^{(1)}_m = \hbox{dim } W^{(1)}_m$.
We have checked (5.11) up to $m=4$ for $D_4$ and 
$m=3$ for $D_5$. 
\par\noindent
{\bf 5.4 Eigenvalues $\Lambda^{(r-1)}_1(u)$ and $\Lambda^{(r)}_1(u)$.}
\hskip0.3cm
Now the relevant auxiliary spaces are
\par\noindent
$W^{(r-1)}_1 \simeq V(\omega_{r-1})$ and
$W^{(r)}_1 \simeq V(\omega_r)$ as $D_r$-modules.
They are the two spin representations, whose weights are all
multiplicity-free and given by (4.21) for 
$V(\omega_{r-1})$ if $\mu_1\mu_2 \cdots \mu_r = -$ and
for 
$V(\omega_r)$ if $\mu_1\mu_2 \cdots \mu_r = +$.
As in the $B_r$ case we shall build the boxes
$\Flect(0.4cm,2.5cm,{\mu_1,\mu_2,\ldots,\mu_r})$
by which the DVF can be written as
$$\eqalignno{
\Lambda^{(r-1)}_1(u) &=\sum_{\{\mu_j = \pm ; \prod_{j=1}^r \mu_j = -\}} 
\overbrace{\Flect(0.4cm,2.5cm,{\mu_1,\mu_2, \cdots , \mu_r})}^{r}_p,
&(5.12{\rm a})\cr
\Lambda^{(r)}_1(u) &=\sum_{\{\mu_j = \pm ; \prod_{j=1}^r \mu_j = +\}} 
\overbrace{\Flect(0.4cm,2.5cm,{\mu_1,\mu_2, \cdots , \mu_r})}^{r}_p.
&(5.12{\rm b})\cr}
$$
We let ${\cal T}^{(r-1)}_1$ and 
${\cal T}^{(r)}_1$ denote the sets of
$\hbox{dim } W^{(r-1)}_1 = \hbox{dim } W^{(r)}_1 = 2^{r-1}$ boxes
in (5.12a) and (5.12b), respectively.
The indices $r$ and 
$p \in \{1,2,\ldots,r-2,+,-\}$ signify the rank of $D_r$
and the quantum space 
$\otimes_{j=1}^N W^{(p)}_1(w_j)$, respectively.
The boxes are again defined by the recursion relations
w.r.t these indices.
By using the operators (4.24), they read,
$$\eqalignno{
\hbox{for } 1 \le p \le r-2,&\cr
\overbrace{\Flect(0.4cm, 1.5cm, {+,+,\xi })}^{r}_p &=
\phi(u+r+p-1)
\tau^Q \overbrace{\Flect(0.4cm, 1.2cm, {+,\xi })}^{r-1}_{p-1},
&(5.13{\rm a})\cr
\overbrace{\Flect(0.4cm, 1.5cm, {+,-,\xi })}^{r}_p &=
\phi(u+r+p-1)
{Q_1(u+r-3)\over Q_1(u+r-1)}
\tau^Q \overbrace{\Flect(0.4cm, 1.2cm, {-,\xi })}^{r-1}_{p-1},
&(5.13{\rm b})\cr
\overbrace{\Flect(0.4cm, 1.5cm, {-,+,\xi })}^{r}_p &=
\phi(u+r-p-1)
{Q_1(u+r+1)\over Q_1(u+r-1)}
\tau^u_2 \tau^Q \overbrace{\Flect(0.4cm, 1.2cm, {+,\xi })}^{r-1}_{p-1},
&(5.13{\rm c})\cr
\overbrace{\Flect(0.4cm, 1.5cm, {-,-,\xi })}^{r}_p &=
\phi(u+r-p-1)
\tau^u_2 \tau^Q \overbrace{\Flect(0.4cm, 1.2cm, {-,\xi })}^{r-1}_{p-1},
&(5.13{\rm d})\cr
\hbox{for } p = \pm,\qquad \quad&\cr
\overbrace{\Flect(0.4cm, 1.5cm, {+,+,\xi })}^{r}_p &=
\phi(u+2r)
\tau^Q \overbrace{\Flect(0.4cm, 1.2cm, {+,\xi })}^{r-1}_{p},
&(5.13{\rm e})\cr
\overbrace{\Flect(0.4cm, 1.5cm, {+,-,\xi })}^{r}_p &=
\phi(u+2r)
{Q_1(u+r-3)\over Q_1(u+r-1)}
\tau^Q \overbrace{\Flect(0.4cm, 1.2cm, {-,\xi })}^{r-1}_{p},
&(5.13{\rm f})\cr
\overbrace{\Flect(0.4cm, 1.5cm, {-,+,\xi })}^{r}_p &=
\phi(u-2)
{Q_1(u+r+1)\over Q_1(u+r-1)}
\tau^u_2 \tau^Q \overbrace{\Flect(0.4cm, 1.2cm, {+,\xi })}^{r-1}_{p},
&(5.13{\rm g})\cr
\overbrace{\Flect(0.4cm, 1.5cm, {-,-,\xi })}^{r}_p &=
\phi(u-2)
\tau^u_2 \tau^Q \overbrace{\Flect(0.4cm, 1.2cm, {-,\xi })}^{r-1}_{p}.
&(5.13{\rm h})\cr
}$$ 
Here $\xi$ denotes arbitrary sequence of $\pm$ symbols with length
$r-2$.
The recursions (5.13) involve both boxes in
${\cal T}^{(r-1)}_1$ and ${\cal T}^{(r)}_1$ and hold for $r \ge 5$.
As in $B_r$ case, we formally 
consider boxes with $p=0$ and fix them
by (4.26b) and the convention explained after it.
We are yet to specify the initial condition, i.e.,
data for $D_4$ case.
As for the dress parts, 
they are given by
$$\eqalign{
dr \Flect(0.4cm, 1.6cm, {+,+,+,+}) &={{Q_4(u-1)}\over{Q_4(u+1)}},\cr
dr \Flect(0.4cm, 1.6cm, {+,+,-,-}) &=
  {{Q_2(u)Q_4(u+3)}\over{Q_2(u+2)Q_4(u+1)}},\cr
dr \Flect(0.4cm, 1.6cm, {+,-,+,-}) &=
  {{Q_1(u+1)Q_2(u+4)Q_3(u+1)}\over{Q_1(u+3)Q_2(u+2)Q_3(u+3)}},\cr
dr \Flect(0.4cm, 1.6cm, {+,-,-,+}) &=
  {{Q_1(u+1)Q_3(u+5)}\over{Q_1(u+3)Q_3(u+3)}},\cr
dr \Flect(0.4cm, 1.6cm, {-,+,+,-}) &=
  {{Q_1(u+5)Q_3(u+1)}\over{Q_1(u+3)Q_3(u+3)}},\cr
dr \Flect(0.4cm, 1.6cm, {-,+,-,+}) &=
  {{Q_1(u+5)Q_2(u+2)Q_3(u+5)}\over{Q_1(u+3)Q_2(u+4)Q_3(u+3)}},\cr
dr \Flect(0.4cm, 1.6cm, {-,-,+,+}) &=
  {{Q_2(u+6)Q_4(u+3)}\over{Q_2(u+4)Q_4(u+5)}},\cr
dr \Flect(0.4cm, 1.6cm, {-,-,-,-}) &={{Q_4(u+7)}\over{Q_4(u+5)}}. \cr
}\eqno(5.14{\rm a})$$
The other 8 are deduced from the above by
$$
dr \Flect(0.4cm,2.4cm,{\mu_1,\mu_2,\mu_3,\mu_4}) = 
dr \Flect(0.4cm,2.4cm,{\mu_1,\mu_2,\mu_3,-\mu_4})
\vert_{Q_3(u) \leftrightarrow Q_4(u)},
\eqno(5.14{\rm b})
$$ 
which is consistent with the diagram automorphism symmetry.
In fact, through the recursions (5.13), the property (5.14b) 
leads to
$$ 
dr \overbrace{\Flect(0.4cm,2.7cm,{\mu_1,\ldots,\mu_{r-1},\mu_r})}^r_p
=
dr \overbrace{\Flect(0.4cm,3.0cm,{\mu_1,\ldots,\mu_{r-1},-\mu_r})}^r_p
\vert_{Q_{r-1}(u) \leftrightarrow Q_r(u)}.
$$
As for the vacuum parts, we shall give their 
general form that includes the initial condition ($r=4$) 
and fulfills the recursions (5.13).
$$\eqalignno{
&vac \overbrace{\Flect(0.4cm,2.0cm,{\mu_1,\ldots,\mu_r})}^r_p
= \cases{
\psi^{(p,r)}_n(u)& for $1 \le p \le r-2$\cr
\psi^{(p,r)}_{n,\mu_r}(u)& for $p = \pm$\cr},
&(5.15{\rm a})\cr
&n = \cases{
\sharp \{ j \mid \mu_j = -, 1 \le j \le p \}
& for $1 \le p \le r-2$\cr
\sharp \{ j \mid \mu_j = -, 1 \le j \le r-1 \}
& for $p = \pm$\cr},&(5.15{\rm b})\cr
&\psi^{(p,r)}_n(u) = 
\prod_{\scriptstyle j=0 \atop 
       \scriptstyle j \neq n}^p \phi(u+r-p+2j-1),
&(5.15{\rm c})\cr
&\psi^{(+,r)}_{n,+}(u) = 
\psi^{(-,r)}_{n,-}(u) = 
\prod_{\scriptstyle j=0 \atop
       \scriptstyle j \neq n+1}^{r+1} \phi(u+2j-2),
&(5.15{\rm d})\cr
&\psi^{(+,r)}_{n,-}(u) = 
\psi^{(-,r)}_{n,+}(u) = \phi(u+2n)
\prod_{\scriptstyle j=0 \atop
       \scriptstyle j \neq n, n+2}^{r+1} \phi(u+2j-2).
&(5.15{\rm e})\cr}
$$
By the definition, $n$ ranges over $0 \le n \le p$ 
in (5.15c) and $0 \le n \le r-1$ in (5.15d,e).
This completes the characterization of 
all the $2^r$ boxes hence the DVF (5.12)
for any $r \ge 4, p \in \{0,1,\ldots, r-2, +, - \}$.
In the rational case ($q \rightarrow 1$) with $p=1$,
a similar recursive description is available in [5].
\par
Let us list a few features explained in section 2.4.
Firstly, the top term (2.12) corresponds to
$$\eqalign{
dr \overbrace{\Flect(0.4cm,2.4cm,{+,\ldots,+,-})}^r_p &= 
{Q_{r-1}(u-1)\over Q_{r-1}(u+1)},\cr
dr \overbrace{\Flect(0.4cm,2.4cm,{+,\ldots,+,+})}^r_p &= 
{Q_r(u-1)\over Q_r(u+1)},\cr}
\eqno(5.16)
$$
where the lhs' are indeed associated with the highest weights
$\omega_{r-1}$ and $\omega_r$ in view of (5.3) and (4.21). 
Secondly, the crossing symmetry (2.18,19) holds.
$$\eqalign{
\tau^C_{2r-2} 
\overbrace{\Flect(0.4cm,2.0cm,{\mu_1,\ldots,\mu_r})}^r_p
&= \overbrace{\Flect(0.4cm,2.3cm,{-\mu_1,\ldots,-\mu_r})}^r_p 
\quad \hbox{ for } 1 \le p \le r-2,\cr
\tau^C_{2r-2} 
\overbrace{\Flect(0.4cm,2.0cm,{\mu_1,\ldots,\mu_r})}^r_p
&= \overbrace{\Flect(0.4cm,2.3cm,{-\mu_1,\ldots,-\mu_r})}^r_{-p} 
\quad \hbox{ for } p = \pm.\cr}
\eqno(5.17)
$$
Thirdly, the coupling rule (2.14a) is valid due to
\proclaim Lemma 5.4.1.
For $1 \le a \le r-1$ the factor $1/Q_a$
is contained in the box
$\overbrace{\Flect(0.4cm,2.0cm,{\mu_1,\ldots,\mu_r})}^r_p$
if and only if $(\mu_a,\mu_{a+1}) = (+,-)$ or $(-,+)$.
Any two such boxes
$\overbrace{\Flect(0.4cm,2.0cm,{\eta,+,-,\xi})}^r_p$ and 
$\overbrace{\Flect(0.4cm,2.0cm,{\eta,-,+,\xi})}^r_p$
share a common color $a$ pole $1/Q_a(u+y)$ for some $y$.
The factor $1/Q_r$ 
is contained in the box
$\overbrace{\Flect(0.4cm,2.0cm,{\mu_1,\ldots,\mu_r})}^r_p$
if and only if $\mu_{r-1} = \mu_r$.
Any two such boxes
$\overbrace{\Flect(0.4cm,1.5cm,{\zeta,+,+})}^r_p$ and 
$\overbrace{\Flect(0.4cm,1.5cm,{\zeta,-,-})}^r_p$ 
share a common color $r$ pole $1/Q_r(u+z)$ for some $z$.
\par
As introduced in the 
beginning of section 4.6, 
let $\hbox{BAE}^r_{p=0}$ be (2.7) with the lhs being always $-1$
Under the BAE, 
the pair of the coupled boxes yield zero residue in total.
We claim this in
\proclaim Theorem 5.4.2.
For $1 \le a \le r-1$, 
let $\eta, \xi$ and $\zeta$ be any $\pm$ sequences with
lengths $a-1, r-a-1$ and $r-2$, respectively.
If the $\hbox{BAE}^r_p$ (2.7) (with $s=1$) 
for $0 \le p \le r-2$ and (5.1) for $p=\pm$ are valid, then
$$\eqalignno{
&Res_{u=-y+iu^{(a)}_k}\Bigl(
\overbrace{\Flect(0.4cm,2.0cm,{\eta,+,-,\xi})}^r_p +
\overbrace{\Flect(0.4cm,2.0cm,{\eta,-,+,\xi})}^r_p \Bigr) = 0,
&(5.18{\rm a})\cr
&Res_{u=-z+iu^{(r)}_k}\Bigl(
\overbrace{\Flect(0.4cm,2.0cm,{\zeta,+,+})}^r_p +
\overbrace{\Flect(0.4cm,2.0cm,{\zeta,-,-})}^r_p \Bigr) = 0,
&(5.18{\rm b})\cr
}$$
where $y$ and $z$ are those in Lemma 5.4.1.
\par
The proof is similar to that for Theorem 4.6.3.
In particular (2.14b) and (2.15) can be shown,
therefore the character limit (2.16) is valid for 
the DVFs (5.12).
(When $p = \pm$, one modifies the $\omega^{(p)}_{s=1}$ in (2.16)
suitably.) 
Notice that both of
the coupled boxes in (5.18) belong to the 
same set ${\cal T}^{(r-1)}_1$ or ${\cal T}^{(r)}_1$.
Thus Lemma 5.4.1 and Theorem 5.4.2 lead to 
\proclaim Theorem 5.4.3.
For $r \ge 4$ and $p \in \{0,1,\ldots, r-2,+,-\}$, 
$\Lambda^{(r-1)}_1(u)$ and 
$\Lambda^{(r)}_1(u)$ in (5.12)
are free of poles provided that the $\hbox{BAE}^r_p$ (2.7) 
(with $s=1$) for 
$0 \le p \le r-2$ and 
(5.1) for $p = \pm$ are valid.
\par
\par\noindent
{\bf 5.5 Relations between two kinds of boxes.}\hskip0.3cm
The elementary boxes
$\Fsquare(0.4cm,a)$ and $\Flect(0.4cm,1.8cm,{\mu_1,\ldots,\mu_r})$
introduced in section 5.1 and 5.4 are related by
\proclaim Theorem 5.5.1.
For $1 \le a \le r-2, k, n, l \in {\bf Z}_{\ge 0}$ such that 
$k+2n+l=a$, take any integers
$1 \le i_1 < \cdots < i_k \le r$ and 
$1 \le j_1 < \cdots < j_l \le r$.
Then the following equality holds between the elements of
${\cal T}^{(a)}_1$ and 
${\cal T}^{(r-1)}_1 \cup {\cal T}^{(r)}_1$ defined in 
section 5.2 and (5.13-15), respectively.
$$
\raise 13ex \hbox{${1 \over F^{(p,r)}_a(u)}$}
\,
\hbox{
   \normalbaselines\m@th\baselineskip0pt\offinterlineskip
   \vbox{ 
      \hbox{$\Fsquare(0.5cm,\hbox{$i_1$})$}\vskip-0.4pt
      \hbox{$\vnaka$}\vskip-0.4pt
	     \hbox{$\Fsquare(0.5cm,\hbox{$i_k$})$}\vskip-0.4pt
      \hbox{$\Fsquare(0.5cm,\hbox{$\overline{r}$})$}\vskip-0.4pt
      \hbox{$\vnaka$}\vskip-0.4pt
	     \hbox{$\Fsquare(0.5cm,\hbox{$r$})$}\vskip-0.4pt
      \hbox{$\Fsquare(0.5cm,\hbox{$\overline{j}_{\ell}$})$}\vskip-0.4pt
      \hbox{$\vnaka$}\vskip-0.4pt
      \hbox{$\Fsquare(0.5cm,\hbox{$\overline{j}_1$})$} 
        }
      }
\,\, 
 \raise 13ex 
  \hbox{$= \bigl(\, \tau^u_{-r+a+1}
        \overbrace{\Flect(0.4cm,1.8cm,{\mu_1,\ldots,\mu_r})}^r_p \, \bigr)
        \bigl(\, \tau^u_{r-a-1}
        \overbrace{\Flect(0.4cm,1.8cm,{\nu_1,\ldots,\nu_r})}^r_p \, \bigr),
    $}\eqno(5.19)
$$
where there are $n$
$\hbox{
   \normalbaselines\m@th\baselineskip0pt\offinterlineskip
   \vbox{ 
      \hbox{$\Fsquare(0.4cm,\hbox{$\overline{r}$})$}\vskip-0.4pt
	     \hbox{$\Fsquare(0.4cm,\hbox{$r$})$}\vskip-0.4pt
        }
      }
$'s in the lhs and $F^{(p,r)}_a$ is defined in (5.9b).
The $\pm$ sequences $\mu$ and $\nu$ in the rhs are determined 
by (4.40b).
\par
Put 
$a \equiv r + \sigma$  mod  2 where
$\sigma = 0$ or $1$.
Then the tableaux in the rhs of (5.19) belong to the following sets.
$$
\overbrace{\Flect(0.4cm,1.0cm,\mu)}^r_p \in 
\cases{ {\cal T}^{(r-\sigma)}_1 &  $l$ even \cr
        {\cal T}^{(r-1+\sigma)}_1 & $l$ odd \cr},\quad
\overbrace{\Flect(0.4cm,1.0cm,\nu)}^r_p \in 
\cases{ {\cal T}^{(r)}_1 &  $l$ even \cr
        {\cal T}^{(r-1)}_1 & $l$ odd \cr}.
\eqno(5.20)
$$
One can rewrite the rhs of (5.19) so as to interchange 
the parity of $l$ in (5.20).
Given any $\pm$ sequences
$\mu = (\mu_1,\ldots,\mu_r)$ and 
$\nu = (\nu_1,\ldots,\nu_r)$, we set
$$
\eqalignno{
e_k(\mu,\nu) &= \sharp \{ j \mid 1 \le j \le k, \mu_j = - \}
- \sharp \{j \mid 1 \le j \le k, \nu_j = - \},
&(5.21{\rm a})\cr
d_y(\mu,\nu) &= \hbox{min }\Bigl(
\{ \infty \} \cup \{ k \mid 1 \le k \le r-1, 
e_k(\mu,\nu) = y \} \Bigr).
&(5.21{\rm b})\cr
}$$
Then we have
\proclaim Lemma 5.5.2.
For any $1 \le a \le r-2$ 
and any $\pm$ sequences
$\mu = (\mu_1,\ldots, \mu_r), \nu = (\nu_1,\ldots, \nu_r)$,
one has
$$\eqalign{
       &\bigl(\, \tau^u_{-r+a+1}
        \overbrace{\Flect(0.4cm,1.8cm,{\mu_1,\ldots,\mu_r})}^r_p \, \bigr)
        \bigl(\, \tau^u_{r-a-1}
        \overbrace{\Flect(0.4cm,1.8cm,{\nu_1,\ldots,\nu_r})}^r_p \, \bigr)
\cr
& =        \bigl(\, \tau^u_{-r+a+1}
\overbrace{\Flect(0.4cm,1.8cm,{\mu^\prime_1,\ldots,\mu^\prime_r})
}^r_p \, \bigr)
        \bigl(\, \tau^u_{r-a-1}
\overbrace{\Flect(0.4cm,1.8cm,{\nu^\prime_1,\ldots,\nu^\prime_r})
}^r_p \, \bigr),\cr}\eqno(5.22{\rm a})
$$
where $\mu^\prime_j$ and $\nu^\prime_j$ are determined by
$$
(\mu^\prime_j,\nu^\prime_j) = \cases{
(\mu_j,\nu_j) & if $1 \le j \le d_{r-a-1}(\mu,\nu)$\cr
(\nu_j,\mu_j) & otherwise \cr}.\eqno(5.22{\rm b})
$$
\par
The Lemma enables the interchange of those $\mu_j$ and $\nu_j$
with $j > d_{r-a-1}(\mu,\nu)$ in the products (5.22a).
In case $d_{r-a-1}(\mu,\nu) = \infty$,
the assertion is trivial.
One may apply Lemma 5.5.2 to rewrite the rhs of (5.19).
A little inspection tells that $1 \le d_{r-a-1}(\mu,\nu) \le r-1$
for any those $\mu$ and $\nu$ appearing there.
Moreover, for such $d = d_{r-a-1}(\mu,\nu)$ one can 
evaluate the difference
$$
\sharp \{j \mid d < j \le r, \mu_j = - \} -
\sharp \{j \mid d < j \le r, \nu_j = - \} = 2n+1
\in 2{\bf Z} + 1,
$$
in terms of the $n$ in Theorem 5.5.1.
Thus Lemma 5.5.2 expresses the rhs of (5.19) 
by the tableaux such that
$$
\overbrace{\Flect(0.4cm,1.0cm,\mu^\prime)}^r_p \in 
\cases{ {\cal T}^{(r-1+\sigma)}_1 &  $l$ even \cr
        {\cal T}^{(r-\sigma)}_1 & $l$ odd \cr},\quad
\overbrace{\Flect(0.4cm,1.0cm,\nu^\prime)}^r_p \in 
\cases{ {\cal T}^{(r-1)}_1 &  $l$ even \cr
        {\cal T}^{(r)}_1 & $l$ odd \cr},
\eqno(5.23)
$$
which is opposite to (5.20).
Based on these observations, 
we can give a similar argument to section 4.7 
that backgrounds Theorem 5.5.1.
There is a degeneracy 
point $u=-2(r-a-1)$ of the $U_q(D^{(1)}_r)$ quantum $R$-matrix [32]
where it yields embedding 
$$\eqalign{
W^{(a)}_1(u) \hookrightarrow 
&W^{(r-1)}_1(u+r-a-1) \otimes W^{(r-1+\sigma)}_1(u-r+a+1),\cr
W^{(a)}_1(u) \hookrightarrow 
&W^{(r)}_1(u+r-a-1) \otimes W^{(r-\sigma)}_1(u-r+a+1),\cr}
\eqno(5.24)
$$
According to [8], (5.24) implies the functional relations
$$\eqalignno{
T^{(r-1)}_1(u+r-a-1)T^{(r-1+\sigma)}_1(u-r+a+1) 
&= T^{(a)}_1(u) + T^\prime(u),&(5.25{\rm a})\cr
T^{(r)}_1(u+r-a-1)T^{(r-\sigma)}_1(u-r+a+1)
&= T^{(a)}_1(u) + T^{\prime \prime}(u),&(5.25{\rm b})\cr}
$$
where $T^\prime(u)$ and $T^{\prime \prime}(u)$
are some matrices commuting with 
all $T^{(b)}_m(v)$'s.
In particular if $a=r-2$ ($\sigma = 0$), 
(5.25) is the last equation in (2.5c)
with $m=1$, hence
$T^\prime(u) = T^{(r-1)}_2(u)$ and 
$T^{\prime \prime}(u) = T^{(r)}_2(u)$.
One may regard (5.25a,b) as equations on the eigenvalues 
and substitute (5.9a) and (5.12).
Then Theorem 5.5.1 tells how one can pick up the 
DVF for $\Lambda^{(a)}_1(u)$ from the lhs.
For example in (5.25a), one depicts the terms in 
$\Lambda^{(a)}_1(u)$ as the lhs of (5.19).
Then the $l$ odd terms are indeed contained in 
$\Lambda^{(r-1)}_1(u+r-a-1)\Lambda^{(r-1+\sigma)}_1(u-r+a+1)$
due to (5.19) and (5.20).
The $l$ even terms 
can also be found by expressing the above product
in terms of the tableaux in (5.23).
\beginsection 6. Discussions

\noindent{\bf 6.1 Summary.}\hskip0.3cm
In this paper we have constructed the 
dressed vacuum forms (DVFs) (2.9b) for 
several eigenvalues $\Lambda^{(a)}_m(u)$
of the row-to-row transfer matrices $T^{(a)}_m(u)$ (1.2) via 
the analytic Bethe ansatz.
Relevant vertex models are those associated 
with the fusion quantum $R$-matrices
$R_{W^{(a)}_m, W^{(p)}_s}(u)$ for $Y(X_r)$ or 
$U_q(X^{(1)}_r)$ with $X_r = B_r, C_r$ and $D_r$.
We have determined the DVFs for all the transfer matrices
$T^{(a)}_1(u)\, (1 \le a \le r)$ 
associated with the fundamental representations
$W^{(a)}_1$ in the sense of [28].
In particular, they have been proved pole-free under
the Bethe ansatz equation, a crucial
property in the analytic Bethe ansatz.
Once the DVFs of $\Lambda^{(a)}_1(u)$ are fixed,
those for the other eigenvalues are
uniquely determined from the $T$-system [8], 
a set of functional relations  
among the transfer matrices.
Based on this we have conjectured the DVFs for several 
$\Lambda^{(a)}_m(u)$ with higher $m$. 
These results extend earlier ones in
[5,10,11,19,20,21].
\par
The DVFs for $\Lambda^{(a)}_m(u)$ 
are Yang-Baxterizations of the characters
of the auxiliary spaces $W^{(a)}_m$.
We have found that they are described by remarkably simple rules
using analogues of the
semi-standard Young tableaux.
We believe that the sets of tableaux
${\cal T}^{(a)}_m$ introduced in this paper
are natural objects that label the base of 
the irreducible finite dimensional modules $W^{(a)}_m$
over the Yangians or the quantum affine algebras.
\par\noindent
{\bf 6.2 Further extensions.}\hskip0.3cm
Let us indicate further applications of our approach.
As can be observed 
through sections 2 to 5, the hypotheses called
the top term (2.12) and the coupling rule (2.14), (2.15) severely
restrict possible DVFs.
This is especially significant when as many 
weight spaces as possible are multiplicity-free (2.13)
in the auxiliary space. 
An interesting example of such a situation is
Yangian analogue of the adjoint representation.
Below we exclude the case $X_r = A_r$, where
the DVF for general eigenvalues is already available [19].
Then it is known [12,28] that 
the Yangian $Y(X_r)$ admits the irreducible representation
$W_{adj}$ isomorphic to
$V(\theta) \oplus V(0)$ as an $X_r$-module.
Here $\theta$ denotes the highest root hence
$V(\theta)$ means the adjoint representation of $X_r$.
One can identify $W_{adj}$
in the family $\{ W^{(a)}_m \}$ by $\theta$ and
the data in appendix A of [33].
$$
(\theta, W_{adj}) = \cases{
(\omega_1, W^{(1)}_1) & $E_7, E_8, F_4, G_2$\cr
(2\omega_1, W^{(1)}_2) & $C_r$\cr
(\omega_2, W^{(2)}_1) & $B_r, D_r$\cr
(\omega_6, W^{(6)}_1) & $E_6$\cr}.
$$
Thus the cases $X_r = B_r, C_r$ and $D_r$ are already covered in this paper.
For $G_2$, the DVF of $\Lambda^{(1)}_1(u)$ has been obtained
recently [11].
Let us turn to the remaining cases,
$\Lambda^{(1)}_1(u)$ of $E_{7,8}, F_4$ and
$\Lambda^{(6)}_1(u)$ of $E_6$.
By the definition, 
$\hbox{dim } W_{adj} = \hbox{dim } X_r + 1$.
All the weights 
in $W_{adj}$ are multiplicity-free
except the null one,  
$\hbox{mult}_0 W_{adj} = r + 1$.
Thus one may try to apply
the top term (2.12), the coupling rule (2.14,15) and 
the crossing symmetry (2.18) 
to possibly determine the $\hbox{dim } X_r - r$ terms
in the DVF corresponding to the root vectors.
We have checked that this certainly works consistently and fix 
those terms uniquely.
Moreover, we have found that
pole-freeness under the BAE requires
exactly $r + 1$ more terms
that make the null weight contribution $(r+1)q^0$ 
in the character limit (2.16).
These features are equally valid in the trigonometric case as well.
Thus the resulting DVFs are candidates of 
the transfer matrix eigenvalues for 
the trigonometric vertex models associated with 
$U_q(E_8^{(1)})$, etc.
The details will appear elsewhere.
It still remains to understand the hypotheses 
(2.12), (2.14) and (2.18) intrinsically and thus to unveil the full
aspects of the analytic Bethe ansatz.
\beginsection Acknowledgments

The authors thank E. Date, A.N. Kirillov, 
M. Kashiwara, T. Nakashima and M. Okado for discussions.
They also thank T. Nakashima for allowing them to use the 
\TeX macros produced by K. Nakahara.
A part of this work is done in Department of Mathematics,
Kyushu University.
\beginsection References

\item{[1]}{Bethe,H.A.: Zur Theorie der Metalle, I.
Eigenwerte und Eigenfunktionen der
linearen Atomkette. Z.Physik {\bf 71} 205-226 (1931)}
\item{[2]}{Yang,C.N.,Yang,C.P.: Thermodynamics of a one-dimensional
system of bosons with repulsive delta-function interaction.
J.Math.Phys.{\bf 10} 1115-1122 (1969) }
\item{[3]}{Sklyanin,E.K.,Takhtajan,L.A., Faddeev,L.D.:
Quantum inverse problem method I.
Theor.Math.Phys.{\bf 40} 688-706 (1980);
Kulish,P.P.,Sklyanin,E.K.: Quantum}
\item{}{spectral transform method. 
Recent developments. Lect.Note.Phys. {\bf 151} 61-119 Springer
1982}
\item{[4]}{Baxter,R.J.: Partition function of the 
eight-vertex lattice model. Ann.Phys.{\bf 70} 193-228 (1972)}
\item{[5]}{Reshetikhin,N.Yu.: The functional equation
method in the theory of exactly soluble quantum systems.
 Sov.Phys.JETP {\bf 57} 691-696 (1983)}
\item{[6]}{Sklyanin,E.K.: Quantum inverse scattering method. 
Selected topics. in Quantum Group and Quantum
Integrable Systems. ed. Mo-Lin Ge   
Singapore: World Scientific 1992}
\item{[7]}{Paz,R.: Feynman's office: The last blackboards.
Phys. Today {\bf 42} 88 (1989)}
\item{[8]}{Kuniba,A.,Nakanishi,T.,Suzuki,J.:
Functional relations
in solvable lattice models I: Functional relations and representation
theory. (hep-th.9307137)}
\item{}{Int.J.Mod.Phys. in press}
\item{[9]}{Kuniba,A.,Nakanishi,T.,Suzuki,J.: Functional relations
in solvable lattice models II: Applications. (hep-th.9310060) 
Int.J.Mod.Phys. in press}
\item{[10]}{Kuniba,A.: Analytic Bethe ansatz and $T$-system in 
$C^{(1)}_2$ vertex models. J.Phys.}
\item{}{A: Math.Gen.{\bf 27} L113-L118 (1994)}
\item{[11]}{Suzuki,J.: Fusion $U_q(G^{(1)}_2)$ vertex models
and analytic Bethe ans{\" a}tze. preprint 
(hep-th.9405201) (1994)}
\item{[12]}{Drinfel'd,V.G.: Hopf algebras and 
quantum Yang-Baxter equation. Sov.Math.}
\item{}{Dokl.{\bf 32} 254-258 (1985)}
\item{[13]}{Drinfel'd,V.G.: Quantum groups. ICM proceedings,
Berkeley, 798-820 (1986)}
\item{[14]}{Jimbo,M.: A $q$-difference analogue of 
$U(g)$ and the Yang-Baxter equation. Lett.}
\item{}{Math.Phys.{\bf 10} 63-69 (1985)}
\item{[15]}{Kac,V.G.: Infinite Dimensional Lie Algebras.
Cambridge: Cambridge University Press 1990}
\item{[16]}{Kirillov,A.N., N.Yu.Reshetikhin,N.Yu.:
Representations of Yangians and multiplicity 
of occurrence of the irreducible components 
of the tensor product of representations of simple Lie algebras.
J.Sov.Math.{\bf 52} (1990) 3156-3164}
\item{[17]}{Baxter,R.J.: Exactly Solved Models in Statistical 
Mechanics. London: Academic 1982}
\item{[18]}{Kirillov,A.N., N.Yu.Reshetikhin,N.Yu.: 
Exact solution of the integrable $XXZ$ Heisenberg model with
arbitrary spin: I. The ground state and the excitation
spectrum. J.Phys.A: Math.Gen.
{\bf 20} 1565-1585 (1987)}
\item{[19]}{Bazhanov,V.V., N.Yu.Reshetikhin,N.Yu.: 
Restricted solid on solid models connected with
simply laced algebras and conformal field theory. J.Phys.A:Math.}
\item{}{Gen. {\bf 23} 1477-1492 (1990)}
\item{[20]}{Reshetikhin,N.Yu.: 
Integrable models of quantum one-dimensional 
magnets with $O(n)$ and $Sp(2k)$ symmetry.
Theor.Math.Phys.{\bf 63} 55-569 (1985)}
\item{[21]}{Reshetikhin,N.Yu.:
The spectrum of the transfer matrices connected with
Kac-Moody algebras. Lett.Math.Phys.{\bf 14} 235-246 (1987)}
\item{[22]}{Kl{\" u}mper,A., P.A.Pearce,P.A.: 
Conformal weights of RSOS lattice models and
their fusion hierarchies. Physica {\bf A183}
304-350 (1992)}
\item{[23]}{Jimbo,N., Kuniba,A., Miwa,T., Okado,M.:
The $A^{(1)}_n$ face models. Commun.}
\item{}{Math. Phys. {\bf 119} 543-565 (1988)}
\item{[24]}{Kashiwara,M., Nakashima,T.: Crystal graphs for 
representations of the}
\item{}{$q$-analogue of classical Lie algebras.
RIMS preprint {\bf 767} (1991)}
\item{[25]}{Nakashima,T.: 
Crystal base and a generalization of the 
Littlewood-Richardson rule for the classical Lie algebras.
Commun.Math.Phys.{\bf 154} 215-243 (1993)}
\item{[26]}{Ogievetsky,E.I., Wiegmann,P.B.:
Factorized $S$-matrix and the Bethe ansatz 
for simple Lie groups. Phys. Lett. {\bf B168}
360-366 (1986)}
\item{[27]}{Drinfel'd,V.G.: A new realization of Yangians and 
quantum affine algebras. Sov.Math.Dokl.{\bf 36} 212-216 (1988)}
\item{[28]}{Chari,V., Pressley,A.:
Fundamental representations of Yangians and
singularities of $R$-matrices. J.reine angew.Math.{\bf 417}
87-128 (1991)}
\item{[29]}{Kuniba,A., Nakanishi,T.:
Spectra in conformal field theories from the Rodgers dilogarithm.
Mod. Phys. Lett.{\bf A7} 
3487-3494 (1992)}
\item{[30]}{Kirillov,A.N.: Identities for the Rodgers dilogarithm
function connected with simple Lie algebras. 
J.Sov.Math.{\bf 47} 2450-2459 (1989)}
\item{[31]}{Kuniba,A., Nakanishi,T., Suzuki,J.:
Characters in conformal field theories from 
thermodynamic Bethe ansatz. Mod. Phys. Lett.
{\bf A8} 1649-1659 (1993)}
\item{[32]}{Reshetikhin,N.Yu.:
Algebraic Bethe ansatz for $SO(n)$ invariant transfer-matirces.
Zapiski nauch. LOMI {\bf 169} 122-140 (1984) (in Russian);
Okado,M.: Quantum $R$ matrices related to
the spin representation of $B_r$ and $D_r$.
Commun. Math. Phys. {\bf 134} 467-486 (1990)}
\item{[33]}{Kuniba,A.: Thermodynamics of the $U_q(X^{(1)}_r)$
Bethe ansatz system with $q$ a root of unity.
Nucl. Phys. {\bf B389} 209-244 (1993)}
\bye